\documentclass[aps,pra,twocolumn,twoside,a4paper,10pt,amsmath,amssymb,superscriptaddress]{revtex4-1}

\usepackage{mathrsfs}
\usepackage{graphicx}
\usepackage{amsmath}
\usepackage{amssymb}
\usepackage{amsfonts}
\usepackage{color}
\usepackage{CJK}
\usepackage{leftidx}

\def\bra#1{\left\langle{#1}\right|}
\def\ket#1{\left|{#1}\right\rangle}
\def\braket#1#2{\left\langle{{#1}}\mathrel{\left|{\vphantom{{#1}{#2}}}\right.\kern-\nulldelimiterspace}{{#2}}\right\rangle}

\begin{document}

%%%%%%%%%%%%%%%%%%%%%%%%%%%%%%%%%%%%%%%%%%%%%%%%%%%%%%%%%%%%%%%%%%%%%%%%%%%%%%%%%
%%%%%%%%%%%%%%%%%%%%%%%%%%%%%%%%%%%%%%%%%%%%%%%%%%%%%%%%%%%%%%%%%%%%%%%%%%%%%

\title{Simultaneous measurement of DC and AC magnetic fields at the Heisenberg limit}
\author{Min Zhuang}
\affiliation{Guangdong Provincial Key Laboratory of Quantum Metrology and Sensing $\&$ School of Physics and Astronomy, Sun Yat-Sen University (Zhuhai Campus), Zhuhai 519082, China}
\affiliation{State Key Laboratory of Optoelectronic Materials and Technologies, Sun Yat-Sen University (Guangzhou Campus), Guangzhou 510275, China}

\author{Jiahao Huang}
\altaffiliation{Email: hjiahao@mail2.sysu.edu.cn, eqjiahao@gmail.com}
\affiliation{Guangdong Provincial Key Laboratory of Quantum Metrology and Sensing $\&$ School of Physics and Astronomy, Sun Yat-Sen University (Zhuhai Campus), Zhuhai 519082, China}

\author{Chaohong Lee}
\altaffiliation{Email: lichaoh2@mail.sysu.edu.cn, chleecn@gmail.com}
\affiliation{Guangdong Provincial Key Laboratory of Quantum Metrology and Sensing $\&$ School of Physics and Astronomy, Sun Yat-Sen University (Zhuhai Campus), Zhuhai 519082, China}
\affiliation{State Key Laboratory of Optoelectronic Materials and Technologies, Sun Yat-Sen University (Guangzhou Campus), Guangzhou 510275, China}

\begin{abstract}
  High-precision magnetic field measurement is an ubiquitous issue in physics and a critical task in metrology.
  Generally, magnetic field has DC and AC components and it is hard to extract both DC and AC components simultaneously.
  The conventional Ramsey interferometry can easily measure DC magnetic fields, while it becomes invalid for AC magnetic fields since the accumulated phases may average to zero.
  Here, we propose a scheme for simultaneous measurement of DC and AC magnetic fields by combining Ramsey interferometry and rapid periodic pulses.
  In our scheme, the interrogation stage is divided into two signal accumulation processes linked by a unitary operation.
  In the first process, only DC component contributes to the accumulated phase.
  In the second process, by applying multiple rapid periodic $\pi$ pulses, only the AC component gives rise to the accumulated phase.
  By selecting suitable input states and the unitary operations in interrogation and readout stages, and the DC and AC components can be extracted by population measurements.
  In particular, if the input state is a GHZ state and two interaction-based operations are applied during the interferometry, the measurement precisions of both DC and AC components can  simultaneously approach the Heisenberg limit.
  Our scheme provides a feasible way to achieve Heisenberg-limited simultaneous measurement of DC and AC fields.
\end{abstract}
\date{\today}

\maketitle

%%%%%%%%%%%%%%%%%%%%%%%%%%%%%%%%%%%%%%%%%%%%%%%%%%%%%%%%%%%%%%%%%%%%%%%%%%%%%%%%%%%%%%%%%%%%%%%%%%%%%%%%%%%%%%%%%%%%%%%%%%%%%%%%%%%%%%%%
\section{Introduction\label{Sec1}}
%%%%%%%%%%%%%%%%%%%%%%%%%%%%%%%%%%%%%%%%%%%%%%%%%%%%%%%%%%%%%%%%%%%%%%%%%%%%%%%%%%%%%%%%%%%%%%%%%%%%%%%%%%%%%%%%%%%%%%%%%%%%%%%%%%%%%%%%%%%%%%%%%%%%%%%%%%%%%%%%%%%%%%%%%%%%
The high-precision measurement of weak magnetic fields is an important problem in diverse areas ranging from fundamental physics~\cite{CWHelstrom1976,SLBraunstein1994,VGiovannetti2006,BMEscher2011,RDemkowiczDobrz2012,CLDegen2017,Vengalattore2007} and material science to geographic metrology and biomedical sensing~\cite{CCTsuei2000,KKobayashi2003,HJMamin2003,DRugar2004,WWasilewski2010}.
Utilizing the well-developed Ramsey techniques, the DC magnetic fields can be detected with ultra-high sensitivity.
While for AC magnetic field measurement, only using Ramsey techniques becomes invalid and various methods of modulation should be employed.
Dynamical decoupling (DD) method, originated for protecting qubits from decoherence, is one of the effective methods for detecting alternating signals~\cite{GdeLange2010,WJKuo2011,LJiang2011,PZanardi2008,PCMaurer2012,HStrobel2014,MSkotiniotis2015,Hosten2016,JGBohnet2016,ILovchinsky2016,SChoi2017,Biercuk2009,Hirose2012,JMBossl2017}.
For example, using a single $\pi$-pulse (spin-echo) or a multi-$\pi$-pulse sequence in the interrogation process in nitrogen-vacancy-based experiments~\cite{JRMaze2008,GBalasubramanian2008,GdeLange2011}, the AC magnetic fields can be effectively detected with high sensitivity.
So far, most studies on magnetic field measurement focus only on DC or AC component, which belongs to a single-parameter estimation problem.
However, in practical scenarios, the magnetic field may have both DC and AC components.
Therefore, simultaneously estimating the DC and AC magnetic fields becomes a challenge~\cite{CWHelstrom1976,Helstrom1967,Paris2009,PCHumphreys2013,AdvPhysX2016,TBaumgratz2016,Proctor2018,MGessner2018,Zhuang2018,Ragy2016}.

On the other hand, it is well known that multi-particle quantum entanglement can offer a significant enhancement of measurement precision~\cite{VGiovannetti2004,VGiovannetti2011,JHuang2014,JGBohnet2016}.
For $N$ individual particles, according to the central limit theorem, the measurement precision scales as the standard quantum limit (SQL), i.e., $\propto 1/\sqrt{N}$.
However, the SQL can be surpassed by using entangled particles.
For an example, by using the Greenberger-Horne-Zeilinger (GHZ) state, the measurement precision can be improved to the Heisenberg-limited scaling, i.e., $\propto 1/N$~\cite{JJBollinger1996,TMonz2011,JHuang2015,SDHuver2008,BLu2019,Lee2006,CLee2012}.
Quantum-enhanced magnetometers have been proposed and realized in various systems, including nitrogen-vacancy defect centers~\cite{FJelezko2004,JMTaylor2008,SKolkowitz2012},
Bose-Einstein condensates~\cite{IMSavukov2005,Vengalattore2007,HXing2016,EDavis2016,TMacri2016,FFrowis2016,Szigeti2017,Nolan2016,JHuang2018}, trapped ions~\cite{SKotler2011,OHosten2016}, solid-state spin systems~\cite{MPackard1954,LRondin2012,RSchirhagl2014,Troiani2018}.

Recently, a protocol about how to perform Floquet enhanced measurements of an AC magnetic field in Ising-interacting spin systems is presented~\cite{arXiv180100042}.
In this scheme, a multi-$\pi$-pulse sequence is applied and the Heisenberg-scaled measurement precision of AC magnetic field is demonstrated by preparing the GHZ state via adiabatic driving~\cite{Lee2006,Zhang2013,Luo2017,Huang032116,YQZou2018,Huang2018,Zou2018}.
However, this scheme requires a single-particle resolved detection.

It is natural to ask: (i) Can one combine Ramsey interferometry and DD method to estimate DC and AC magnetic fields simultaneously? (ii) Can the measurement precisions simultaneously surpass the SQL or even attain the Heisenberg limit by employing quantum many-body entanglement?
(iii) If the Heisenberg-limited measurements are available, can the realization be accomplished without single-particle resolved detection?
In this article, we propose a scheme for estimating DC and AC magnetic fields simultaneously by combing Ramsey interferometry and periodic modulation.
Our scheme contains three stages: initialization, interrogation, and readout.
In particular, the interrogation process is divided into two signal accumulation processes and a unitary operation.
In the first signal accumulation process, no operations is applied and only the DC component is imprinted onto the accumulated phase.
In the second signal accumulation process, a periodic $\pi$-pulse sequence is applied, and only the AC component contributes to the accumulated phase.
By extracting the total accumulated phase, the DC and AC components can be inferred respectively.
We find that, if the initial state is prepared as a GHZ state, and applying suitable interaction-based operations in interrogation and readout stages~\cite{EDavis2016,TMacri2016,FFrowis2016,Szigeti2017,Nolan2016,JHuang2018,Mirkhalaf2018,Anders2018,Burd2019,Linnemann2016}, both the measurement precisions of DC and AC components can exhibit the Heisenberg-limited scaling simultaneously.
Our scheme may open up a feasible way for measuring DC and AC magnetic fields simultaneously at the Heisenberg limit.

This paper is organized as follows.
In Sec.~\ref{Sec2}, we introduce our scheme on simultaneous measurement of DC and AC magnetic fields.
In Sec.~\ref{Sec3}, within our scheme, we study three different interferometry processes in detail with individual particles as well as entangled particles.
The measurement precisions of DC and AC magnetic fields via three different interferometry processes are analytically obtained.
In Sec.~\ref{Sec4}, we discuss the experimental feasibility of our scheme.
%
%More importantly, input a GHZ state and applied two suitable interaction-based operation in the interferometry process, the measurement precisions of the two parameters can attain the Heisenberg limit simultaneously.
%
Finally, we give a brief summary in Sec.~\ref{Sec5}.
%%%%%%%%%%%%%%%%%%%%%%%%%%%%%%%%%%%%%%%%%%%%%%%%%%%%%%%%%%%%%%%%%%%%%%%%%%%%%%%%%%%%%%%%%%%%%%%%%%%%%%%%%%%%%%%%%%%%%%%%%%%%%%%%%%%%%%%%
\section{General scheme \label{Sec2}}
Our protocol of simultaneous measurement of DC and AC magnetic fields is presented below.
We consider an ensemble of two-mode bosonic system with $N$ particles coupled to an external magnetic field ${\textbf{B}}(t)= \left[B_{1} +B_{2}\sin(\omega_{s}t)\right]{\textbf{z}}$ oscillating along the $z$-direction.
Here, $\omega_s$ corresponds to the oscillation frequency (which is assumed known). $B_1$ and $B_2$ respectively stand for the strengths of DC and AC components to be measured.
The two modes can be suitably selected as two magnetic levels, and hereafter we label them as spins $\ket{\uparrow}$ and $\ket{\downarrow}$, respectively.
The $N$ spin-1/2 bosonic system can be well characterized by the collective spin operators:
$\hat{J}_{x}=\frac{1}{2}(\hat{a}^{\dag}\hat{b}+\hat{a}\hat{b}^{\dag}),\hat{J}_{y}=\frac{1}{2i}(\hat{a}^{\dag}\hat{b}-\hat{a}\hat{b}^{\dag}),
\hat{J}_{z}=\frac{1}{2}(\hat{a}^{\dag}\hat{a}-\hat{b}^{\dag}\hat{b})$,
where $\hat{a}$ and $\hat{b}$ denote annihilation operators for spins $\ket{\uparrow}$ and $\ket{\downarrow}$, respectively.
The system state can be represented in terms of Dike basis $\{|J,m_{k}\rangle\}$ with $\hat{J}_{k}|J,m_{k}\rangle=m_{k}|J,m_{k}\rangle({k}\in\{x,y,z\})$, $J=\frac{N}{2}$ and $m_k = -J,-J + 1, ..., J+1, J$.
Thus, the Hamiltonian describing the system coupled to the external magnetic field ${\textbf{B}}(t)$ can be expressed as
\begin{equation}\label{Eq:HamS}
\hat{H}_{B}(t)={\textbf{B}}(t) \cdot \textbf{J}=\left[B_{1}+B_{2}\sin(\omega_{s}t)\right]\hat{J}_{z}.
\end{equation}
Our goal is to measure the two parameters $B_{1}$ and $B_{2}$ simultaneously.

\begin{figure}[!htp]
 \includegraphics[width=1\columnwidth]{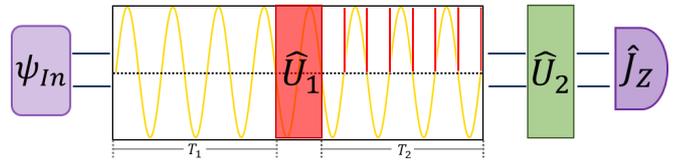}
  \caption{\label{Fig1}(color online).
  Protocol of simultaneous measurement of DC and AC magnetic fields. The scheme consists of three stages: initialization, interrogation, and readout.
  First, the system is initialized in a desired input state.
  Then, the interrogation stage is divided into two signal accumulation processes and a unitary operation $\hat{U}_{1}$.
  In the former accumulation process, the information of DC component can be encoded in the accumulated phase $\phi_{DC}$.
  While for the latter accumulation process, a sequence of $\pi$ pulses (denoted by thin red lines) are applied at every node of the external magnetic field (denoted by yellow line) to imprint the information of AC component into $\phi_{AC}$.
  Finally, an operation $\hat{U}_{2}$ is performed for recombination before the readout stage, and the two parameters can be extracted from half-population difference measurement.}
\end{figure}

The scheme on simultaneous measurement of $B_{1}$ and $B_{2}$ can be divided into three stages: (i) initialization, (ii) interrogation, and (iii) readout, see Fig.~\ref{Fig1}.
Throughout this paper, we assume all time-evolution processes are unitary and set $\hbar = 1$.
In the initialization stage, a suitable input state $| \Psi \rangle_{\textrm{In}}$ is prepared.
Then, the input state undergoes an interrogation stage for signal accumulation.
At the start of this stage, the system interacts with the magnetic field for a duration $T_1$ and a unitary operation $\hat U_1$ is performed.
Then, a multi-$\pi$-pulse sequence is successionally applied at every node of the magnetic field. The periodic multi-pulse sequence is locked resonantly to the AC component with frequency $2\omega_s$, and the system evolves for a duration $T_2$.
In the final readout stage, another unitary operation $\hat U_2$ is applied for recombination, and the half-population difference measurement is implemented to extract the information of the parameters $B_1$ and $B_2$.

The key for simultaneous measurement of DC and AC magnetic fields is the interrogation stage.
At this stage, there are two signal accumulation processes.

For the first signal accumulation process, the system is exposed under the magnetic field for a duration $T_1$ without any operations.
Here, the duration $T_1$ should be properly chosen as $T_{1}={2\pi}n/{{\omega}_{s}}$ with $n$ being an integer.
In this case, the AC component will have no contribution to the accumulated phase since $\int_0^{T_1} B_2 \sin(\omega_s t) dt=0$.
On the contrary, the DC component will give rise to an accumulated phase $\phi_{DC}$ proportional to $B_1 T_1$.

For the second signal accumulation process, a multi-$\pi$-pulse sequence is applied.
The $\pi$ pulses (i.e., $\hat{\textrm{P}}_{\pi}=e^{-i \pi \hat{J}_{x}}$) flip the spin states ($\ket{\uparrow}\rightarrow\ket{\downarrow}$ and $\ket{\downarrow}\rightarrow\ket{\uparrow}$).
The effect of a $\pi$ pulse corresponds to the transformation $\hat{J}_{z}$ $\rightarrow$ $-\hat{J}_{z}$ onto the instantaneous Hamiltonian.
Thus, when the $\pi$ pulses are applied at every node of the magnetic field, the effect of the AC component will lead to an accumulated phase $\phi_{AC}$.
The periodic multi-$\pi$-pulse sequence will effectively give rise to a non-zero time-averaged signal strength ${2B_2}/{\pi}$ for each half-cycle, and the accumulated phase $\phi_{AC}$ is proportional to ${2B_2 T_2}/{\pi}$ if $T_{2}={\pi}k/{{\omega}_{s}}$ with $k$ an integer.
%$\frac{\omega_s}{\pi}\int_0^{\pi/\omega_s} B_2 \sin(\omega_s t)dt = 2B_2/\pi$

%
At the same time, the multi-$\pi$-pulse sequence will cancel out the influences of the DC component, which is similar with the principle of spin echoes.
In this way, the strength of the AC component can be encoded in the accumulated phase $\phi_{AC}$.
%
%In total, the whole accumulated phase $\phi$ after the interrogation stage have two parts, i.e., $\phi=\phi_{DC}+\phi_{AC}$.

One can also use effective Hamiltonians to describe the signal accumulation stage~\cite{Kuwahara2016, Abanin2017}.
For the former one (duration $T_1$), the time-averaged magnetic field comes from the DC component, and the effective quasi-static Hamiltonian reads,
\begin{equation}\label{HamE1}
	\hat{H}_{B}^{\text{eff1}} = B_1 \hat J_z.
\end{equation}
For the latter one (duration $T_2$), the time-averaged magnetic field comes from the AC component, and the effective quasi-static Hamiltonian becomes,
\begin{equation}\label{HamE2}
	\hat{H}_{B}^{\text{eff2}} = \frac{2B_2}{\pi} \hat J_z.
\end{equation}

In order to distinguish the two accumulated phases $\phi_{DC}$ and $\phi_{AC}$, one needs to rotate the state in the middle of the interrogation stage.
The selection of unitary operation $\hat U_1$ depends on the input state and will have influences on the final measurement precisions, which will be discussed in the next section.

The output state after the interrogation stage can be expressed in the form of $\ket{\Psi_{\textrm{Out}}}\!=\!e^{-i\hat{H}_{B}^{\text{eff2}}T_{2}}\hat{U}_{1} e^{-i\hat{H}_{B}^{\text{eff1}}T_{1}}\ket{\Psi_{\textrm{In}}}$ .
Finally, in the readout stage, another unitary operation $\hat{U}_{2}$ is performed on $\ket{\Psi}_{\textrm{Out}}$ for recombination.
Thus, the final state before half-population difference measurement can be written as
\begin{equation}\label{Eq:final_state}
	\ket{\Psi_{\textrm{final}}}=\hat{U}_{2}e^{-i\hat{H}_{B}^{\text{eff2}}T_{2}}\hat{U}_{1} e^{-i\hat{H}_{B}^{\text{eff1}}T_{1}} \ket{\Psi_{\textrm{In}}}.
\end{equation}
The final state contains the information of the estimated DC and AC magnetic field strengths $B_1$ and $B_2$.

%In practice, the two parameters $B_1$ and $B_2$ are in the unit of hertz.
%
%In our calculation, without loss of generality, $B_1$ and $B_2$ are set dimensionless.
%
%Besides, the durations $T_1$ and $T_2$ are set to be the same for convenience, i.e., $T_1=T_2=T$.
%
%
According to the multiparameter quantum estimation theory~\cite{CWHelstrom1976,Paris2009,AdvPhysX2016}, the precision of the two parameters $B_1$ and $B_2$ can be determined according to the covariance matrix $\mathrm{Cov}(B_{1},B_{2})$, which is bounded by
\begin{equation}\label{Eq:Cov-FQ}
  \mathrm{Cov}(B_{1},B_{2})\geq \left[\mathnormal{\mathbf{F}}_C(B_{1},B_{2})\right]^{-1}\geq \left[\mathnormal{\mathbf{F}}_Q(B_{1},B_{2})\right]^{-1},
\end{equation}
with $\mathnormal{\mathbf{F}}_C(B_{1},B_{2})$ and $\mathnormal{\mathbf{F}}_Q(B_{1},B_{2})$ being the classical Fisher information matrix (CFIM) and quantum Fisher information matrix (QFIM), respectively.
The Fisher information matrix provides an asymptotic measure of the amount of information on the parameters of a system.

Since the variance of the two parameters are the diagonal terms of the covariance matrix $\mathrm{Cov}(B_{1},B_{2})$, they satisfy the inequalities
\begin{eqnarray}\label{InEq:Delta_B1_B2_general}
 \Delta^2{B_{k}}\geq[\mathnormal{\mathbf{F}}_C(B_{1},B_{2})]^{-1}_{kk}\geq[\mathnormal{\mathbf{F}}_Q(B_{1},B_{2})]^{-1}_{kk},
\end{eqnarray}
where $k=1,2$.
According to inequalities~\eqref{InEq:Delta_B1_B2_general}, the elements of CFIM and QFIM determine the classical Cram\'{e}r-Rao bound (CCRB) and the quantum Cram\'{e}r-Rao bound (QCRB) for simultaneous measurement of $B_1$ and $B_2$.
The detailed calculations for CFIM and QFIM are shown in Appendix A.

More practically, particularly in experiments, one need find a suitable observable to approach the theoretical precision bounds.
According to the quantum estimation theory, the measurement precisions of the estimated parameters can be given by the error propagation formula,
\begin{equation}\label{Eq:Parameter uncertainty}
\Delta B_{k}=\frac{(\Delta{\hat{J}_{z}})_{\text{f}}}{|\partial{\langle\hat{J}_{z}\rangle_{\text{f}}}/ \partial{B_{k}}|}, \quad k=1,2.
\end{equation}
Here, $(\Delta{\hat{J}_{z}})_{\text{f}}$ and $\langle\hat{J}_{z}\rangle_{\text{f}}$ are respectively the standard deviation and expectation of $\hat{J}_z$ in the form of
\begin{equation}\label{Eq:Deviation}
(\Delta{\hat{J}_{z}})_{\text{f}}=\sqrt{\langle\hat{J}_z^2\rangle_{\text{f}}-\langle\hat{J}_{z}\rangle_{\text{f}}^2},
\end{equation}
\begin{equation}\label{Eq:Expectation}
\langle\hat{J}_{z}\rangle_{\text{f}}=\!\bra{\Psi_{\text{final}}} \hat{J}_{z} \ket{\Psi_{\text{final}}},
\end{equation}
with
\begin{equation}\label{Eq:Expectation2}
\langle\hat{J}_{z}^2\rangle_{\text{f}}=\!\bra{\Psi_{\text{final}}} \hat{J}_{z}^2 \ket{\Psi_{\text{final}}}.
\end{equation}
In the next section, we will discuss how to realize the Heisenberg-limited simultaneous measurement of $B_1$ and $B_2$ within this framework.
%%%%%%%%%%%%%%%%%%%%%%%%%%%%%%%%%%%%%%%%%%%%%%%%%%%%%%%%%%%%%%%%%%%%%%%%%%%%%%%%%%%%%%%%%%%%%%%%%%%%%%%%%%%%%%%%%%%%%%%%%%%%%%%%%%%%%%%%
\section{Measurement precisions\label{Sec3}}
%%%%%%%%%%%%%%%%%%%%%%%%%%%%%%%%%%%%%%%%%%%%%%%%%%%%%%%%%%%%%%%%%%%%%%%%%%%%%%%%%%%%%%%%%%%%%%%%%%%%%%%%%%%%%%%%%%%%%%%%%%%%%%%%%%%%%%%%
In the following, we illustrate how to simultaneously estimate the two parameters and give the measurement precisions under three scenarios.
For individual particles without entanglement, the measurement precisions for the two parameters can just approach SQL.
For entangled particles in GHZ state, the measurement precision of DC component can attain Heisenberg limit by using interaction-based readout.
Further, if another interaction-based operation is performed in the interrogation stage, the measurement precisions of DC and AC components can both exhibit  Heisenberg scaling simultaneously.

%%%%%%%%%%%%%%%%%%%%%%%%%%%%%%%%%%%%%%%%%%%%%%%%%%%%%%%%%%%%%%%%%%%%%%%%%%%%%%%%%%%%%%%%%%%%%%%%%%%%%%%%%%%%%%%%%%%%%%%%%%%%%%%%%%%%%%%%
\subsection{Individual particles\label{A}}
%%%%%%%%%%%%%%%%%%%%%%%%%%%%%%%%%%%%%%%%%%%%%%%%%%%%%%%%%%%%%%%%%%%%%%%%%%%%%%%%%%%%%%%%%%%%%%%%%%%%%%%%%%%%%%%%%%%%%%%%%%%%%%%%%%%%%%%%
%
We first consider individual particles without any entanglement.
Suppose all the particles are prepared in the spin coherent state (SCS) $\ket{\Psi}_{\textrm{SCS}}=e^{-i\frac{\pi}{2}\hat J_y}\ket{N/2,-N/2}$.
This input state can be easily generated by applying a $\pi/2$ pulse on the state of all particles in spin-down $\ket{\downarrow}$.
In this situation, one can choose $\hat{U}_{1}=\hat{U}_{2}=e^{-i\frac{\pi}{2}{\hat{J}_{y}}}$.
Then, the final state before the half-population difference measurement can be written as
\begin{eqnarray}\label{Evo_CSC}
|\Psi_{\text{final}}\rangle
&&=e^{-i\frac{\pi}{2}{\hat{J}_{y}}} e^{-i\hat{H}_B^{\text{eff2}}T_{2}} e^{-i\frac{\pi}{2}{\hat{J}_{y}}} e^{-i\hat{H}_B^{\text{eff1}}T_{1}}\ket{\Psi}_{\textrm{SCS}}.
\end{eqnarray}
In an explicit form, the final state becomes (See Appendix B for derivation)
\begin{widetext}
\begin{eqnarray}\label{Evo_CSC}
|\Psi_{\text{final}}\rangle
&& =\sum_{m_{z}=\!-J}^{J}\!\!\frac{\sqrt{C_{J}^{m_{z}}}}{2^{J}}\left[\cos(\frac{B_{1}T_{1}}{2})e^{-i \frac{B_{2}T_{2}}{\pi}}\!-\!i\sin(\frac{B_{1}T_{1}}{2})e^{\!i \frac{B_{2}T_{2}}{\pi}}\right]^{J-m_{z}}
\left[\cos(\frac{B_{1}T_{1}}{2})e^{-i \frac{B_{2}T_{2}}{\pi}}\!+\!i\sin(\frac{B_{1}T_{1}}{2})e^{\!i \frac{B_{2}T_{2}}{\pi}}\right]^{J+m_{z}}\!|J,m_{z}\rangle, \nonumber\\
\end{eqnarray}
\end{widetext}
where $C_{J}^{m_{z}}={\frac{(2J)!}{(J+m_{z})!(J-m_{z})!}}$ is the binomial coefficient.
According to Eq.~\eqref{Eq:FQM}, the elements of the QFIM can be written as
\begin{eqnarray}\label{Eq:FQ11_CSS}
[\mathnormal{\mathbf{F}}_Q^{\textrm{{SCS}}}(B_{1},B_{2})]_{11}=NT_{1}^2,
\end{eqnarray}
\begin{eqnarray}\label{Eq:FQ12_CSS}
[\mathnormal{\mathbf{F}}_Q^{\textrm{{SCS}}}(B_{1},B_{2})]_{12}=[\mathnormal{\mathbf{F}}_Q^{\textrm{{SCS}}}(B_{1},B_{2})]_{21}=0,
\end{eqnarray}
\begin{widetext}
\begin{eqnarray}\label{Eq:FQ22_CSS}
&&[\mathnormal{\mathbf{F}}_Q^{\textrm{{SCS}}}\!(B_{1},B_{2})]_{22} = 16\sum_{m_{z}=\!-J}^{J}\!{C_{J}^{m_{z}}}(\frac{m_{z} T_{2}}{\pi})^2\left[\cos(B_{1}T_{1}/2)\right]^{2J+2m_{z}}\!\left[\sin(B_{1}T_{1}/2)\right]^{2J-2m_{z}}\nonumber\\
&&-\!16\!\left\{\!\sum_{m_{z}=\!-J}^{J}\!\!\!\!\frac{{C_{J}^{m_{z}}}m_{z} T_{2}}{{\pi}}\left[\cos(B_{1}T_{1}/2)\right]^{2J+2m_{z}}\!\left[\sin(B_{1}T_{1}/2)\right]^{2J-2m_{z}}\!\!\right\}\!\!
\left\{\sum_{m_{z}=\!-J}^{J}\!\!\frac{{C_{J}^{m_{z}}}m_{z} T_{2}}{\pi}\left[\cos(B_{1}T_{1}/2)\right]^{2J+2m_{z}}\!\left[\sin(B_{1}T_{1}/2)\right]^{2J-2m_{z}}\!\!\right\}.\nonumber\\
\end{eqnarray}
\end{widetext}
For parameter $B_{1}$, the corresponding QCRB $\Delta B_{1}^{\textrm{Q}}=\frac{1}{\sqrt{N}T_1}$, which attains the SQL.
For parameter $B_{2}$, its QCRB is dependent on the parameter $B_{1}$ and $T_{1}$.
When $B_{1}T_{1}=(1/2+k)\pi$, the corresponding QCRB $\Delta B_{2}^{\textrm{Q}}=\frac{\pi}{2\sqrt{N}T_2}$, which scales as the SQL with a constant $\frac{\pi}{2}$.
To obtain the CCRB that saturate the QCRB, we consider the positive operator-valued measure (POVM) in the form of $\hat{\Pi}_{m_z}=\ket{J,m_{z}}\bra{J,m_{z}}$ with $\hat{J_{z}}\ket{J,m_{z}}=m_{z}\ket{J,m_{z}}$ and $\sum_{m_z=-J}^J \hat{\Pi}_{m_z}=\hat{I}$.
%
% The probability associated to the POVM is $P_{m_{z}}=Tr\left[\rho(B_{1},B_{2})\hat{\Pi}_{m_z}\right]$.
%
According to the Eq.~\eqref{Eq:FIM}, we can obtain the CFIM and the CCRB for the two parameters.
For parameter $B_{1}$, the optimal value of CCRB saturate the SQL, as shown in Fig.~\ref{Fig5} (solid circles).
For parameter $B_{2}$, the optimal value of CCRB saturate the corresponding QCRB, as shown in Fig.~\ref{Fig5} (hollow circles).

Further, we consider the measurement precision via practical half-population difference measurement.
After some algebra, the expectations of half-population difference and the square of half-population difference on the final state can also be explicitly written as
\begin{widetext}
\begin{equation}\label{Jz_SCS}
\langle J_{z} \rangle_{\text{f}} = \sum_{m_{z}=-J}^{J}\frac{m_{z} {C_{J}^{m_{z}}}}{4^{J}}\left[1+\sin(B_{1}T_{1})\sin(\frac{2B_{2}T_{2}}{\pi})\right]^{J-m_{z}}\left[1-\sin(B_{1}T_{1})\sin(\frac{2B_{2}T_{2}}{\pi})\right]^{J+m_{z}},
\end{equation}
\begin{equation}\label{Jz2_SCS}
\langle J_{z}^{2} \rangle_{\text{f}} = \sum_{m_{z}=-J}^{J}\frac{m^{2}_{z}{C_{J}^{m_{z}}}}{4^{J}}\left[1+\sin(B_{1}T_{1})\sin(\frac{2B_{2}T_{2}}{\pi})\right]^{J-m_{z}}\left[1-\sin(B_{1}T_{1})\sin(\frac{2B_{2}T_{2}}{\pi})\right]^{J+m_{z}}.
\end{equation}
\end{widetext}
From Eq.~\eqref{Jz_SCS}, it is found that the information of the estimated two parameters $B_1$ and $B_2$ can be inferred from the bi-sinusoidal oscillation of the half-population difference.
In our calculation, the durations $T_1$ and $T_2$ are set to be the same for convenience, i.e., $T_1=T_2$.
Thus, one can obtain the two main oscillation frequencies $\frac{B_{1}}{2}+\frac{B_{2}}{\pi}$ and $\left|\frac{B_{1}}{2}-\frac{B_{2}}{\pi}\right|$ by fast Fourier transform (FFT), and further extract the values of $B_1$ and $B_2$.
However, the oscillations are independent on the total particle number $N$.

\begin{figure}[!htp]
 \includegraphics[width=1\columnwidth]{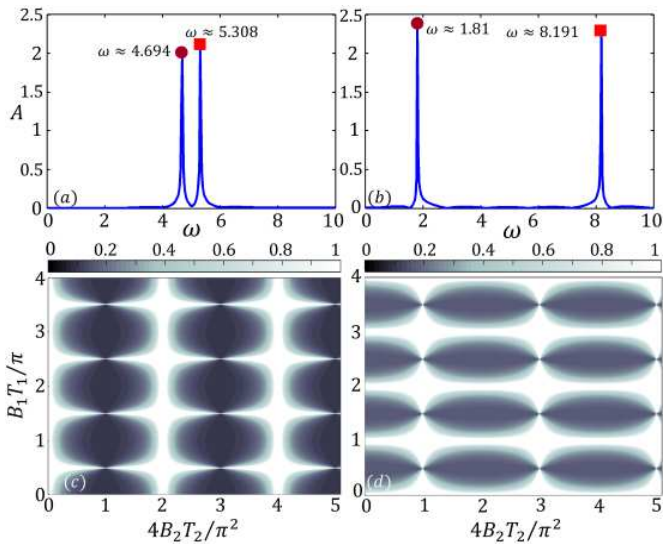}
  \caption{\label{Fig2}(color online).
  FFT spectra and measurement precisions with individual particles under $\hat{U}_{1}=\hat{U}_{2}=e^{-i\frac{\pi}{2}{\hat{J}_{y}}}$.
  The FFT spectra of half-population measurement $\langle J_{z} \rangle_{\text{f}}$ for total particle number $N = 10$ with (a) $B_{1}=10$, $B_{2}=1$ and (b) $B_{1}=10$, $B_{2}=10$. Here, $T_{1}=T_{2}=1$.
  The two main oscillation frequencies are very close to $\frac{B_{1}}{2}+\frac{B_{2}}{\pi}$ and $\left|\frac{B_{1}}{2}-\frac{B_{2}}{\pi}\right|$.
  The variations of measurement precisions (c) $\Delta B_{1}$ and (d) $\Delta B_{2}$ with $B_{1}$ and $B_{2}$ for total particle number $N=100$. }
\end{figure}

In Fig.~\ref{Fig2}~(a) and (b), the FFT spectra for $N=10$ with different $B_{1}$ and $B_{2}$ are shown.
The numerical results perfectly agree with our theoretical predictions.
This imply that the values of $B_{1}$ and $B_{2}$ can be simultaneously obtained only by the half-population difference measurement.

Substituting Eq.~\eqref{Jz_SCS} and Eq.~\eqref{Jz2_SCS} into Eq.~\eqref{Eq:Parameter uncertainty}, one can analytically obtain the measurement precisions $\Delta B_{1}$ and $\Delta B_{2}$.
However, the explicit forms are too cumbersome and therefore we only show the numerical results.
In Fig.~\ref{Fig2}~(c) and (d), how the measurement precisions $\Delta B_{1}$ and $\Delta B_{2}$ change with $B_{1}$ and $B_{2}$ are shown.
%
%Further, we obtain the scaling of minimum measurement precision for $\Delta B_{1}^{\textrm{min}}$ and $\Delta B_{2}^{\textrm{min}}$ versus total particle number $N$, as shown in Fig.~\ref{Fig4}.
%
According to our results, we find that the optimal measurement precision $\Delta B_{1}^{\textrm{min}}$ located at $B_{1}T_{1}=k\pi$ and $B_{2}T_{2}=(2k+1){\pi^2}/{4}$ with $k$ an integer, and
the optimal measurement precision $\Delta B_{2}^{\textrm{min}}$ located at $B_{1}T_{1}=(1/2+k)\pi$ and $B_{2}T_{2}=k{\pi^2}/{2}$.
Since the input state is not entangled, both $\Delta B_{1}^{\textrm{min}}$ and $\Delta B_{2}^{\textrm{min}}$ cannot surpass the SQL as expected.
For parameter $B_{1}$, the optimal measurement precision saturate the SQL.
For parameter $B_{2}$, the optimal measurement precision is a bit worse than the SQL, which is multiplied by a factor $\frac{2}{\pi}$, as shown in Fig.~\ref{Fig6} (circles).

Further, to find out the optimal measurement precision for estimating the two parameters $B_{1}$ and $B_{2}$ simultaneously.
We minimize the sum of the variance $\Delta^2 B_{1}+\Delta^2 B_{2}$, and denote the corresponding measurement precision for the two parameters as $\Delta B_{1}^{\textrm{sim}}$ and $\Delta B_{2}^{\textrm{sim}}$ respectively~\cite{PCHumphreys2013,TBaumgratz2016,Zhuang2018,Ragy2016}.
The minima are located in the vicinity of $B_{1}T_{1}=(1/2+k)\pi$ and $B_{2}T_{2}=(1+2k)\pi^2/{4}$, see Fig.~\ref{Fig7}~(a).
%
%However, the $\Delta B_{1}$ blows up at $B_{1}T_{1}=(1/2+k)\pi$, since the value ${|\partial{\langle\hat{J}_{z}\rangle_{\text{f}}}/ \partial{B_{1}}|}$ approaches to $0$.
%
%The same result for $\Delta B_{2}$ occurs at $B_{2}T_{2}=(1+2k)\pi^2/{4}$, due to the value ${|\partial{\langle\hat{J}_{z}\rangle_{\text{f}}}/ \partial{B_{2}}|}$ approaches to $0$.
%
The measurement precisions $\Delta B_{1}^{\textrm{sim}}$ and $\Delta B_{2}^{\textrm{sim}}$ are both have the SQL scalings but with a constant simultaneously.
According to the fitting results, the measurement precision for $\Delta B_{1}^{\textrm{sim}}\approx \frac{\sqrt{2.5}}{\sqrt{N}T_{1}}$ and $\Delta B_{2}^{\textrm{sim}}\approx \frac{\pi\sqrt{1.75}}{2\sqrt{N}T_{2}}$, as shown in Fig.~\ref{Fig7}~(c) and (d) (circles).
%%%%%%%%%%%%%%%%%%%%%%%%%%%%%%%%%%%%%%%%%%%%%%%%%%%%%%%%%%%%%%%%%%%%%%%%%%%%%%%%%%%%%%%%%%%%%%%%%%%%
\subsection{Entangled particles with one interaction-based operation\label{B}}
%%%%%%%%%%%%%%%%%%%%%%%%%%%%%%%%%%%%%%%%%%%%%%%%%%%%%%%%%%%%%%%%%%%%%%%%%%%%%%%%%%%%%%%%%%%%%%%%%%%%%%%%%%%%%%%%%%%%%%%%%%%%%%%%%%%%%%%%
Entanglement is an effective quantum resource to improve the measurement precision.
For single parameter estimation, by employing GHZ state as the input state, the measurement precision can be improved to the Heisenberg limit.
Here, we try to use an input GHZ state to perform the simultaneous measurement.
We choose a $\pi/2$ pulse $\hat{U}_{1}=e^{-i\frac{\pi}{2}{\hat{J}_{y}}}$ in the interrogation stage and an interaction-based operation $\hat{U}_{2}=e^{i\frac{\pi}{2}{\hat{J}_{z}^2}}e^{-i\frac{\pi}{2}{\hat{J}_{y}}}$ in the readout stage.
The interaction-based readout is a powerful technique for achieving Heisenberg limit via GHZ state without single-particle resolved detection~\cite{EDavis2016,TMacri2016,FFrowis2016,Szigeti2017,Nolan2016,JHuang2018,Mirkhalaf2018,Anders2018}, which is now feasible in experiments~\cite{Burd2019,Hosten2016}.
Therefore, the final state before the half-population difference can be written as
\begin{equation}\label{Evo_GHZ1}
	\ket{\Psi_{\text{final}}} = e^{-i\frac{\pi}{2}{\hat{J}_{y}}} e^{i\frac{\pi}{2}{\hat{J}_{z}^{2}}} e^{-i\hat{H}_B^{\text{eff2}}T_{2}} e^{-i\frac{\pi}{2}{\hat{J}_{y}}} e^{-i\hat{H}_B^{\text{eff1}}T_{1}} \ket{\Psi}_\textrm{GHZ},
\end{equation}
with
\begin{equation}\label{GHZ}
	\ket{\Psi}_\textrm{GHZ}=\frac{1}{\sqrt{2}}(\ket{J,J}+\ket{J,-J}).
\end{equation}
The final state $|\Psi_{\text{final}}\rangle$ has analytical form when $N$ is an even number (see Appendix C for derivation).
\begin{widetext}
For $J=N/2$ is even, it reads
\begin{eqnarray}\label{Evo_GHZ1_even}
|\Psi_{\text{final}}\rangle
&=&\frac{\cos({2B_{1}J T_{1}})-\sin({2B_{1}J T_{1}})}{\sqrt{2}}\sum_{m_{z}=-J}^{J} \sqrt{C_{J}^{m_{z}}}\left[i\cos(\frac{B_{2}T_{2}}{\pi})\right]^{J+m_{z}}\left[\sin(\frac{B_{2}T_{2}}{\pi})\right]^{J-m_{z}} |J,m_{z}\rangle \nonumber\\
&+&\frac{\cos({2B_{1}J T_{1}})+\sin({2B_{1}J T_{1}})}{\sqrt{2}}\sum_{m_{z}=-J}^{J} \sqrt{C_{J}^{m_{z}}}\left[-i\sin(\frac{B_{2}T_{2}}{\pi})\right]^{J+m_{z}}\left[\cos(\frac{B_{2}T_{2}}{\pi})\right]^{J-m_{z}}|J,m_{z}\rangle.
\end{eqnarray}

For $J=N/2$ is odd, it reads
\begin{eqnarray}\label{Evo_GHZ1_odd}
|\Psi_{\text{final}}\rangle
&=&-\frac{\cos({2B_{1}J T_{1}})+\sin({2B_{1}J T_{1}})}{\sqrt{2}}\sum_{m_{z}=-J}^{J} \sqrt{C_{J}^{m_{z}}}\left[i\cos(\frac{B_{2}T_{2}}{\pi})\right]^{J+m_{z}}\left[\sin(\frac{B_{2}T_{2}}{\pi})\right]^{J-m_{z}} |J,m_{z}\rangle \nonumber\\
&+&\frac{\cos({2B_{1}J T_{1}})-\sin({2B_{1}J T_{1}})}{\sqrt{2}}\sum_{m_{z}=-J}^{J} \sqrt{C_{J}^{m_{z}}}\left[-i\sin(\frac{B_{2}T_{2}}{\pi})\right]^{J+m_{z}}\left[\cos(\frac{B_{2}T_{2}}{\pi})\right]^{J-m_{z}} |J,m_{z}\rangle.
\end{eqnarray}
\end{widetext}
For this case, the elements of the QFIM are
\begin{eqnarray}\label{Eq:FQ11_GHZ1}
\left[\mathnormal{\mathbf{F}}_Q^{\textrm{{GHZ1}}}(B_{1},B_{2})\right]_{11}=N^{2}T_{1}^{2},
\end{eqnarray}
\begin{eqnarray}\label{Eq:FQ12_GHZ1}
\left[\mathnormal{\mathbf{F}}_Q^{\textrm{{GHZ1}}}(B_{1},B_{2})\right]_{12}
=\left[\mathnormal{\mathbf{F}}_Q^{\textrm{{GHZ1}}}(B_{1},B_{2})\right]_{21}=0,
\end{eqnarray}
\begin{eqnarray}\label{Eq:FQ22_GHZ1}
\left[\mathnormal{\mathbf{F}}_Q^{\textrm{{GHZ1}}}\!(B_{1},B_{2})\right]_{22}={4 N T_{2}^2}/{\pi^2}.
\end{eqnarray}
For parameter $B_{1}$, its QCRB is $\Delta B_{1}^{\textrm{Q}}=\frac{1}{N T_1}$, which attains the Heisenberg limit.
For parameter $B_{2}$, its QCRB is $\Delta B_{2}^{\textrm{Q}}=\frac{\pi}{2\sqrt{N}T_2}$, which only scales as the SQL.
To obtain the CCRB that saturate the QCRB, we also consider POVM in the form of $\hat{\Pi}_{m_z}=\ket{J,m_{z}}\bra{J,m_{z}}$ with $\sum_{m_z=-J}^J \hat{\Pi}_{m_z}=\hat{I}$.
%
% The probability associated to the POVM is $P_{m_{z}}=Tr\left[\rho(B_{1},B_{2})\hat{\Pi}_{m_z}\right]$.
%
According to the Eq.~\eqref{Eq:FIM}, we can obtain the CFIM and the CCRB for the two parameters.
For parameter $B_{1}$, the optimal value of CCRB saturate the Heisenberg limit, as shown in Fig.~\ref{Fig5} (solid squares).
For parameter $B_{2}$, the optimal value of CCRB saturate the SQL with a constant $\frac{\pi}{2}$, as shown in Fig.~\ref{Fig5} (hollow squares).

Further, we consider the half-population difference measurement.
After the readout stage, the expectation of the half-population difference and the square of half-population difference on the final state can be explicitly written as
\begin{widetext}
\begin{eqnarray}\label{Jz_GHZ1}
\langle J_{z} \rangle_{\text{f}}
&&=\frac{1}{2}(-1)^{J}\cos^{2J}(\frac{B_{2}T_{2}}{\pi})\sin^{2J}(\frac{B_{2}T_{2}}{\pi})
[1-\sin({2B_{1}J T_{1}})]\sum_{m_{z}=-J}^{J}C_{J}^{m_{z}} m_{z}\cot^{2m_{z}}(\frac{B_{2}T_{2}}{\pi})
\nonumber\\
&&+\frac{1}{2}(-1)^{J}\cos^{2J}(\frac{B_{2}T_{2}}{\pi})\sin^{2J}(\frac{B_{2}T_{2}}{\pi})
[1+\sin({2B_{1}J T_{1}})]\sum_{m_{z}=-J}^{J}C_{J}^{m_{z}}m_{z}\tan^{2m_{z}}(\frac{B_{2}T_{2}}{\pi}),
\end{eqnarray}
and
\begin{eqnarray}\label{Jz2_GHZ1}
\langle J_{z}^2 \rangle_{\text{f}}
=&&\frac{1}{2}(-1)^{J}\cos^{2J}(\frac{B_{2}T_{2}}{\pi})\sin^{2J}(\frac{B_{2}T_{2}}{\pi})
[1-\sin({2B_{1}J T_{1}})]\sum_{m_{z}=-J}^{J}C_{J}^{m_{z}}m_{z}^2\cot^{2m_{z}}(\frac{B_{2}T_{2}}{\pi})
\nonumber\\
&&+\frac{1}{2}(-1)^{J}\cos^{2J}(\frac{B_{2}T_{2}}{\pi})\sin^{2J}(\frac{B_{2}T_{2}}{\pi})
[1+\sin({2B_{1}J T_{1}})]\sum_{m_{z}=-J}^{J}C_{J}^{m_{z}}m_{z}^2\tan^{2m_{z}}(\frac{B_{2}T_{2}}{\pi})
\nonumber\\
&&-(-1)^{J}\cos^{2J}(\frac{B_{2}T_{2}}{\pi})\sin^{2J}(\frac{B_{2}T_{2}}{\pi})
\cos({2B_{1}J T_{1}})\sum_{m_{z}=-J}^{J}C_{J}^{m_{z}}m_{z}^2.
\end{eqnarray}
\end{widetext}
%
%Similarly, the analytic expression for $\langle J_{z}^{2} \rangle_{\text{f}}$ can be obtained by replacing $m$ by $m^{2}$ in Eqs.~\eqref{Jz_GHZ1_even} and \eqref{Jz_GHZ1_odd}, and therefore $\Delta \hat J_z$ can also be obtained.

\begin{figure}[!htp]
 \includegraphics[width=1\columnwidth]{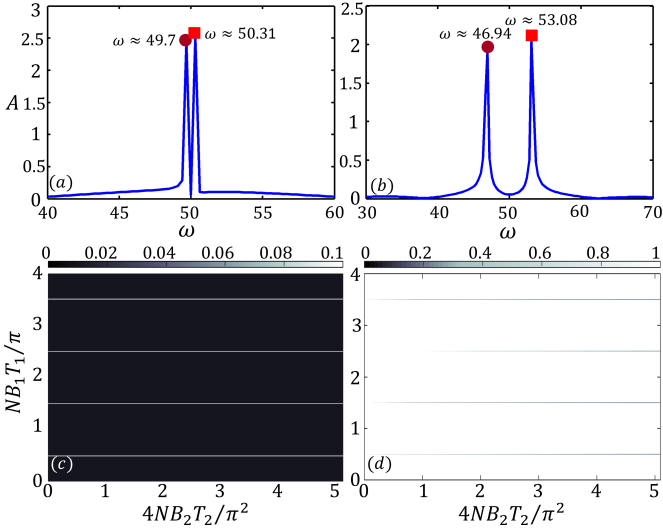}
  \caption{\label{Fig3}(color online).
  FFT spectra and measurement precisions with input GHZ state under $\hat{U}_{1}=e^{-i\frac{\pi}{2}{\hat{J}_{y}}}$ and $\hat{U}_{2}=e^{i\frac{\pi}{2}{\hat{J}_{z}^2}}e^{-i\frac{\pi}{2}{\hat{J}_{y}}}$.
  The FFT spectra of half-population difference measurement $\langle J_{z} \rangle_{\text{f}}$ for total particle number $N = 10$ with (a) $B_{1}=10$, $B_{2}=1$ and (b) $B_{1}=10$, $B_{2}=10$. Here, $T_{1}=T_{2}=1$.
  The two main oscillation frequencies are very close to $\frac{N B_{1}}{2}+\frac{B_{2}}{\pi}$ and $\left|\frac{N B_{1}}{2}-\frac{B_{2}}{\pi}\right|$.
  The variations of measurement precisions (c) $\Delta B_{1}$ and (d) $\Delta B_{2}$ with $B_{1}$ and $B_{2}$ for total particle number $N=100$. }
\end{figure}

By comparison with individual particles, one of the main frequencies of the bi-sinusoidal oscillation of $\langle J_{z}^{2} \rangle_{\text{f}}$ becomes proportional to the total particle number $N=2J$.
Similarly, choosing the durations  $T_1=T_2$ and using FFT analysis, one can get the estimated values of $B_{1}$ and $B_{2}$.
The FFT spectra of $\langle J_{z} \rangle_{\text{f}}$ indicate the two main oscillation frequencies are $\frac{N B_{1}}{2}+\frac{B_{2}}{\pi}$ and $|\frac{N B_{1}}{2}-\frac{B_{2}}{\pi}|$, see Fig.~\ref{Fig3}~(a) and (b) for the FFT results with $N=10$.
This implies that $B_{1}$ and $B_{2}$ can be simultaneously obtained via the half-population difference measurement and the measurement precision of $B_1$ can be improved faster with $N$.

In Fig.~\ref{Fig3}~(c) and (d), we show how the measurement precisions for $\Delta B_{1}$ and $\Delta B_{2}$ vary with $B_{1}$ and $B_{2}$.
The optimal $\Delta B_{1}^{\textrm{min}}$ is located at $NB_{1}T_{1}=k\pi$ while the optimal $\Delta B_{2}^{\textrm{min}}$ is at the position of $NB_{1}T_{1}=k\pi/{2}$.
However, they are both insensitive to $B_{2}$.

To confirm the dependence of $\Delta B_1$ and $\Delta B_2$ on the total particle number $N$, the measurement precision scalings of $\Delta B_{1}^{\textrm{min}}$ and $\Delta B_{2}^{\textrm{min}}$ versus $N$ are shown in Fig.~\ref{Fig6} (squares).
The measurement precision $\Delta B_{1}^{\textrm{min}}$ can attain the Heisenberg limit.
However,  the $\Delta B_{2}^{\textrm{min}}$ is still exhibit the scaling of SQL.

%%%%%%%%%%%%%%%%%%%%%%%%%%%%%%%%%%%%%%%%%%%%%%%%%%%%%%%%%%%%%%%%%%%%%%%%%%%%%%%%%%%%%%%%%%%%%%%%%%%%
\subsection{Entangled particles with two interaction-based operations\label{c}}
%%%%%%%%%%%%%%%%%%%%%%%%%%%%%%%%%%%%%%%%%%%%%%%%%%%%%%%%%%%%%%%%%%%%%%%%%%%%%%%%%%%%%%%%%%%%%%%
\begin{figure}[!htp]
 \includegraphics[width=1\columnwidth]{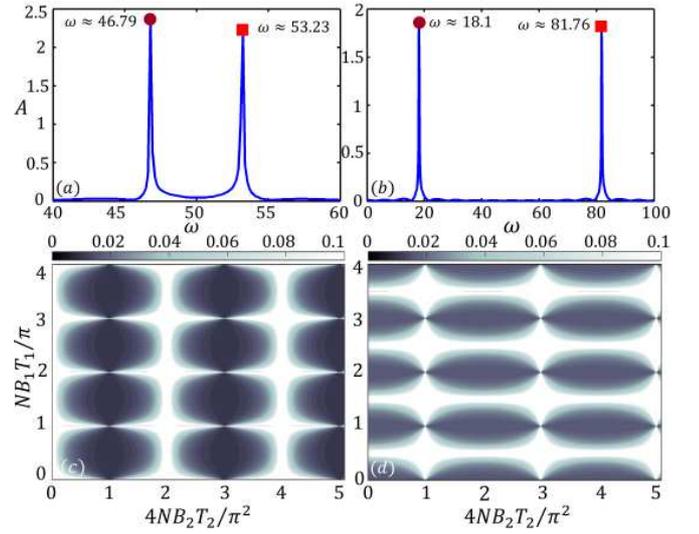}
  \caption{\label{Fig4}(color online).
    FFT spectra and measurement precisions with input GHZ state under $\hat{U}_{1} = e^{i\frac{\pi}{2}\!{\hat{J}_{x}^2}}$ and $\hat{U}_{2} = e^{-i\frac{\pi}{2}{\hat{J}_{y}}}e^{i\frac{\pi}{2}{\hat{J}_{z}^{2}}}e^{-i\frac{\pi}{2}{\hat{J}_{y}}}$.
  The FFT spectra of half-population difference measurement $\langle J_{z} \rangle_{\text{f}}$ for total particle number $N = 10$ with (a) $B_{1}=10$, $B_{2}=1$ and (b) $B_{1}=10$, $B_{2}=10$. Here, $T_{1}=T_{2}=1$.
  The two main oscillation frequencies are very close to $\frac{N B_{1}}{2}+\frac{N B_{2}}{\pi}$ and $\left|\frac{N B_{1}}{2}-\frac{N B_{2}}{\pi}\right|$.
  The variations of measurement precisions (c) $\Delta B_{1}$ and (d) $\Delta B_{2}$ with $B_{1}$ and $B_{2}$ for total particle number $N=100$. }
\end{figure}

Only the measurement precision of DC component can attain the Heisenberg limit is not our ultimate goal.
How to make the measurement precisions of DC and AC magnetic fields simultaneously approach the Heisenberg limit?
We tackle this problem by introducing another interaction-based operation in the middle of the interrogation stage.
That is, we replace $\hat{U}_{1}=e^{-i\frac{\pi}{2}{\hat{J}_{y}}}$ by $\hat{U}_{1}=e^{i\frac{\pi}{2}{\hat{J}_{x}^2}}$.
The unitary operator $\hat{U}_{1}=e^{i\frac{\pi}{2}{\hat{J}_{x}^2}}$ is an interaction-based operation along x-axis.
To cooperate with $\hat{U}_1$, we choose $\hat{U}_{2}=e^{-i\frac{\pi}{2}{\hat{J}_{y}}}e^{i\frac{\pi}{2}{\hat{J}_{z}^{2}}}e^{-i\frac{\pi}{2}{\hat{J}_{y}}}$.
The recombination operation is also an interaction-based operation which comprises a nonlinear dynamics sandwiched by two $\pi/2$ pulses.
By applying this sequence, the final state can be written as
\begin{eqnarray}\label{Evo_GHZ2}
\ket{\Psi_{\text{final}}} \!&=&\! e^{-i\frac{\pi}{2} \hat{J}_{y}} e^{i\frac{\pi}{2}\hat{J}_{z}^{2}} e^{-i\frac{\pi}{2}\hat{J}_{y}} e^{-i \hat{H}_{B}^{\text{eff2}} T_{2}} e^{i\frac{\pi}{2} \hat{J}_{x}^2} e^{-i \hat{H}_{B}^{\text{eff1}} T_{1}} \! \ket{\Psi}_\textrm{GHZ}\nonumber \\
    &=& \text{A}_{1} |J,J\rangle + \text{A}_{2} |J,-J\rangle.
\end{eqnarray}
For $J=N/2$ is even, the two coefficients reads
\begin{eqnarray}\label{Cofficient2}
&&\text{A}_{1} = \frac{1}{\sqrt{2}}\cos(B_{1}JT_{1})\!\left[\cos(\frac{2B_{2}JT_{2}}{\pi})-\sin(\frac{2B_{2}JT_{2}}{\pi})\right] \nonumber\\
&-& \frac{i}{\sqrt{2}}\sin(B_{1}JT_{1})\!\left[\cos(\frac{2B_{2}JT_{2}}{\pi})+\sin(\frac{2B_{2}JT_{2}}{\pi})\right],
\end{eqnarray}
\begin{eqnarray}\label{Cofficient3}
&&\text{A}_{2} = \frac{1}{\sqrt{2}}\cos(B_{1}JT_{1})\!\left[\cos(\frac{2B_{2}JT_{2}}{\pi})+\sin(\frac{2B_{2}JT_{2}}{\pi})\right] \nonumber\\
&+& \frac{i}{\sqrt{2}}\sin(B_{1}JT_{1})\!\left[\cos(\frac{2B_{2}JT_{2}}{\pi})-\sin(\frac{2B_{2}JT_{2}}{\pi})\right].
\end{eqnarray}
For $J=N/2$ is odd, the two coefficients reads
\begin{eqnarray}\label{Cofficient4}
&&\text{A}_{1} = \frac{1}{\sqrt{2}}\cos(B_{1}JT_{1})\!\left[\cos(\frac{2B_{2}JT_{2}}{\pi})+\sin(\frac{2B_{2}JT_{2}}{\pi})\right] \nonumber\\
&+& \frac{i}{\sqrt{2}}\sin(B_{1}JT_{1})\!\left[\sin(\frac{2B_{2}JT_{2}}{\pi})-\cos(\frac{2B_{2}JT_{2}}{\pi})\right],
\end{eqnarray}
\begin{eqnarray}\label{Cofficient5}
&&\text{A}_{2} = \frac{1}{\sqrt{2}}\cos(B_{1}JT_{1})\!\left[\cos(\frac{2B_{2}JT_{2}}{\pi})-\sin(\frac{2B_{2}JT_{2}}{\pi})\right] \nonumber\\
&+& \frac{i}{\sqrt{2}}\sin(B_{1}JT_{1})\!\left[\cos(\frac{2B_{2}JT_{2}}{\pi})+\sin(\frac{2B_{2}JT_{2}}{\pi})\right].
\end{eqnarray}
In this situation, the elements of QFIM are
%[\mathnormal{\mathbf{F}}_Q(B_{1},B_{2})]_{1,1}
\begin{eqnarray}\label{Eq:FQ11_GHZ2}
\left[\mathnormal{\mathbf{F}}_Q^{\textrm{{GHZ2}}}(B_{1},B_{2})\right]_{11}=N^{2}T_{1}^{2},
\end{eqnarray}
\begin{eqnarray}\label{Eq:FQ12_GHZ2}
\left[\mathnormal{\mathbf{F}}_Q^{\textrm{{GHZ2}}}(B_{1},B_{2})\right]_{12}
=\left[\mathnormal{\mathbf{F}}_Q^{\textrm{{GHZ2}}}(B_{1},B_{2})\right]_{21}=0,
\end{eqnarray}
\begin{eqnarray}\label{Eq:FQ22_GHZ2}
\left[\mathnormal{\mathbf{F}}_Q^{\textrm{{GHZ2}}}(B_{1},B_{2})\right]_{22}=4\left[\cos(NB_{1}T_{1})N T_{2}/{\pi}\right]^{2}.
\end{eqnarray}

\begin{figure}[!htp]
 \includegraphics[width=\columnwidth]{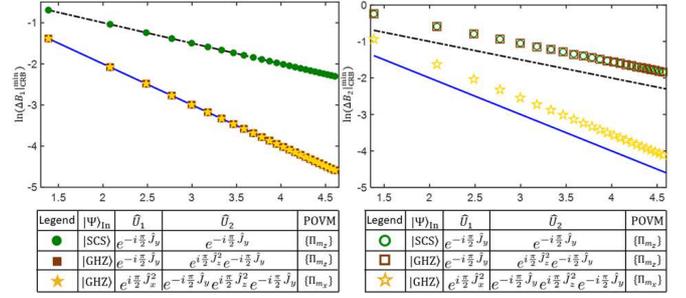}
  \caption{\label{Fig5}(color online).
  The log-log measurement precision scaling via CFIM versus total particle number for (a) DC magnetic field $B_1$ and (b) AC magnetic field $B_2$.
  The circles are the results for SCS with $\hat{U}_{1}=\hat{U}_{2}=e^{-i\frac{\pi}{2}{\hat{J}_{y}}}$ and $\{\hat{\Pi}_{m_z}\}$ ;
  The squares are the results for GHZ state with $\hat{U}_{1}=e^{-i\frac{\pi}{2}{\hat{J}_{y}}}$, $\hat{U}_{2}=e^{i\frac{\pi}{2}{\hat{J}_{z}^2}}e^{-i\frac{\pi}{2}{\hat{J}_{y}}}$ and $\{\hat{\Pi}_{m_z}\}$ ;
  The stars are the results for GHZ state with $\hat{U}_{1}=e^{i\frac{\pi}{2}{\hat{J}_{x}^2}}$, $\hat{U}_{2}\!=\!e^{-i\frac{\pi}{2}{\hat{J}_{y}}}e^{i\frac{\pi}{2}{\hat{J}_{z}^{2}}}e^{-i\frac{\pi}{2}{\hat{J}_{y}}}$ and $\{\hat{\Pi}_{m_x}\}$.
  The black dotted lines denote the SQL, $1/\sqrt{N}$.
  The blue solid lines indicate the Heisenberg limit, $1/{N}$.}
\end{figure}

For parameter $B_{1}$, its QCRB is $\Delta B_{1}^{\textrm{Q}}=\frac{1}{N T_1}$, which attains the Heisenberg limit.
For parameter $B_{2}$, its QCRB is dependent on $B_1$ and $T_1$. The corresponding QCRB $\Delta B_{2}^{\textrm{Q}}=\frac{\pi}{2 N T_2 \cos(N B_1 T_1)}$.
When $N B_1 T_1=k\pi$, $\Delta B_{2}^{\textrm{Q}}=\frac{\pi}{2 N T_2}$ exhibits the Heisenberg-limited scaling.
Unlike the previous two examples, we cannot use the POVM $\hat{\Pi}_{m_z}=\ket{J,m_{z}}\bra{J,m_{z}}$ to saturate the QCRB since the corresponding CFIM becomes non-inverted.
Instead, we choose the POVM in the form of $\hat{\Pi}_{m_x}=\ket{J,m_{x}}\bra{J,m_{x}}$ with $\hat{J}_{x}\ket{J,m_{x}}=m_{x}\ket{J,m_{x}}$ and $\sum_{m_x=-J}^J \hat{\Pi}_{m_x}=\hat{I}$.
According to the Eq.~\eqref{Eq:FIM}, we can obtain the CFIM and the CCRB for the two parameters.
For parameter $B_{1}$, the optimal value of CCRB saturate the Heisenberg limit, as shown in Fig.~\ref{Fig5} (solid stars).
For parameter $B_{2}$, the optimal value of CCRB also have the Heisenberg-limited scaling with only a constant $\frac{\pi}{2}$, as shown in Fig.~\ref{Fig5} (hollow stars).

Further, we study the measurement precision via the half-population difference measurement.
The final half-population difference can be written explicitly as
\begin{equation}\label{Jz_GHZ2}
\langle J_{z} \rangle_{\text{f}}=(-1)^{J+1} J\cos(2B_{1}JT_{1})\sin\left(\frac{4B_{2}JT_{2}}{\pi}\right).
\end{equation}
Clearly, the main frequencies of the bi-sinusoidal oscillation of $\langle J_{z} \rangle_{\text{f}}$ both becomes proportional to $N=2J$.
In the case of $T_1=T_2$, the FFT of $\langle J_{z} \rangle_{\text{f}}$ explicitly indicates that the two main oscillation frequencies are $\frac{N B_{1}}{2}+\frac{NB_{2}}{\pi}$ and $|\frac{N B_{1}}{2}-\frac{NB_{2}}{\pi}|$, as shown in Fig.~\ref{Fig4}~(a) and (b).

\begin{figure}[!htp]
 \includegraphics[width=\columnwidth]{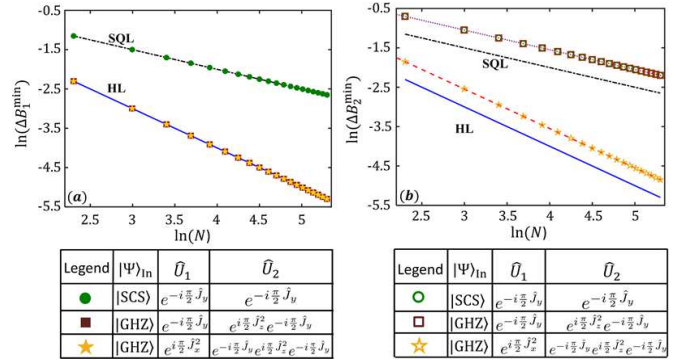}
  \caption{\label{Fig6}(color online).
  The log-log measurement precision scaling versus total particle number for (a) DC magnetic field $B_1$ and (b) AC magnetic field $B_2$.
  The circles are the results for SCS with $\hat{U}_{1}=\hat{U}_{2}=e^{-i\frac{\pi}{2}{\hat{J}_{y}}}$;
  The squares are the results for GHZ state with $\hat{U}_{1}=e^{-i\frac{\pi}{2}{\hat{J}_{y}}}$ and $\hat{U}_{2}=e^{i\frac{\pi}{2}{\hat{J}_{z}^2}}e^{-i\frac{\pi}{2}{\hat{J}_{y}}}$;
  The stars are the results for GHZ state with $\hat{U}_{1}=e^{i\frac{\pi}{2}{\hat{J}_{x}^2}}$ and $\hat{U}_{2}\!=\!e^{-i\frac{\pi}{2}{\hat{J}_{y}}}e^{i\frac{\pi}{2}{\hat{J}_{z}^{2}}}e^{-i\frac{\pi}{2}{\hat{J}_{y}}}$.
  The black dotted lines denote the SQL, $1/\sqrt{N}$.
  The blue solid lines indicate the Heisenberg limit, $1/{N}$.}
\end{figure}

\begin{figure}[!htp]
 \includegraphics[width=\columnwidth]{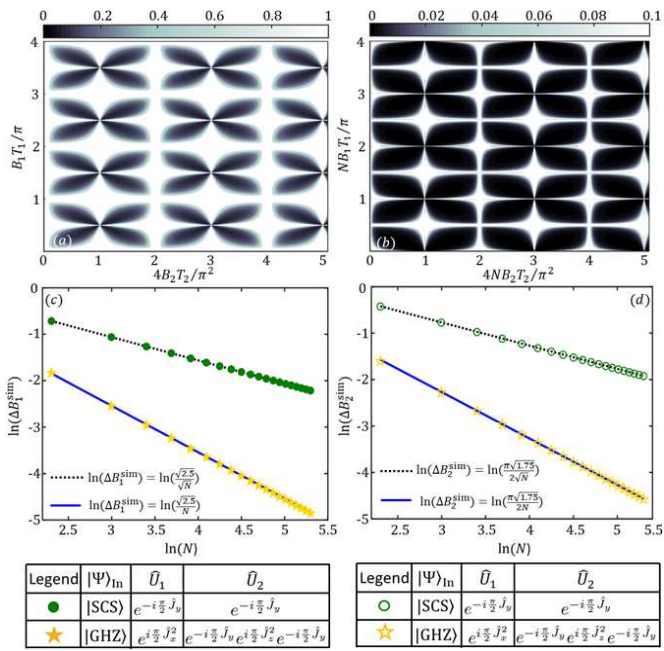}
  \caption{\label{Fig7}(color online).
  The variations of the sum of $\Delta^2 B_1 + \Delta^2 B_2$ versus $B_1$ and $B_2$ for (a) SCS and (b) GHZ state with $\hat{U}_{1}=e^{i\frac{\pi}{2}{\hat{J}_{x}^2}}$ and $\hat{U}_{2}\!=\!e^{-i\frac{\pi}{2}{\hat{J}_{y}}}e^{i\frac{\pi}{2}{\hat{J}_{z}^{2}}}e^{-i\frac{\pi}{2}{\hat{J}_{y}}}$. Here, the total particle number $N=100$.
  The log-log optimal simultaneous measurement precision scaling versus total particle number for (c) DC magnetic field $B_1$ and (d) AC magnetic field $B_2$.
  The circles are the results for SCS with $\hat{U}_{1}=\hat{U}_{2}=e^{-i\frac{\pi}{2}{\hat{J}_{y}}}$, the black dotted lines are the corresponding fitting curves.
  The stars are the results for GHZ state with $\hat{U}_{1}=e^{i\frac{\pi}{2}{\hat{J}_{x}^2}}$ and $\hat{U}_{2}\!=\!e^{-i\frac{\pi}{2}{\hat{J}_{y}}}e^{i\frac{\pi}{2}{\hat{J}_{z}^{2}}}e^{-i\frac{\pi}{2}{\hat{J}_{y}}}$,
  the blue solid lines are the corresponding fitting curves.}
\end{figure}

Moreover, the square of half-population difference is independent on the two parameters, which becomes $\langle J_{z}^{2} \rangle_{\text{f}}={J^{2}}=N^2/4$.
According to Eq.~\eqref{Eq:Parameter uncertainty}, the analytical expression of measurement precisions for $B_{1}$ and $B_{2}$ are
\begin{equation}\label{Delta_B1}
\Delta{B_{1}}=\frac{1}{NT_{1}}\frac{\sqrt{1-\cos^{2}(NB_{1}T_{1})\sin^{2}(\frac{2NB_{2}T_{2}}{\pi})}}
{\left|\sin(NB_{1}T_{1})\sin(\frac{2NB_{2}T_{2}}{\pi})\right|},
\end{equation}
\begin{equation}\label{Delta_B2}
\Delta{B_{2}}=\frac{\pi}{2NT_{2}}\frac{\sqrt{1-\cos^{2}(NB_{1}T_{1})\sin^{2}(\frac{2NB_{2}T_{2}}{\pi})}}
{\left|\cos(NB_{1}T_{1})\cos(\frac{2NB_{2}T_{2}}{\pi})\right|}.
\end{equation}
The optimal measurement precision $\Delta B_{1}^{\textrm{min}}$ can be obtained when $NB_{1}T_{1}=(1/2+k)\pi$ and $NB_{2}T_{2}=(2k+1)\pi^2/{4}$, while the optimal measurement precision $\Delta B_{2}^{\textrm{min}}$ can be obtained when $NB_{1}T_{1}=k\pi$ and $NB_{2}T_{2}=k\pi^2/{2}$.
In Fig.~\ref{Fig4}~(c) and (d), we show how the measurement precisions for $\Delta B_{1}$ and $\Delta B_{2}$ changes with $B_{1}$ and $B_{2}$.
The $\Delta B_{1}^{\textrm{min}}=\frac{1}{NT_{1}}$ attains the Heisenberg limit, while $\Delta B_{2}^{\textrm{min}}=\frac{\pi}{2NT_{2}}$ preserves the Heisenberg scalings but with a constant $\pi/2$.
Our analytical results are confirmed by numerical calculations, see Fig.~\ref{Fig6} (stars).
From Eqs.~\eqref{Delta_B1} and \eqref{Delta_B2}, the  measurement precisions $\Delta B_{1}$ and $\Delta B_{2}$ can exhibit Heisenberg-limited scaling simultaneously.
It is obvious that, when $NB_{1}T_{1}=(1+2k)\pi/4$ and $NB_{2}T_{2}=(1+2k)\pi^2/{8}$, one have $\Delta B_{1}=\frac{\sqrt{3}}{NT_{1}}$ and $\Delta B_{2}=\frac{\sqrt{3}\pi}{2NT_{2}}$ simultaneously.
Further, one can minimize the sum of variance $\Delta^2 B_{1}+\Delta^2 B_{2}$ to find out the optimal simultaneous measurement precision for the two parameter.
The minima are located in the vicinity of $NB_{1}T_{1}=k\pi$ and $NB_{2}T_{2}=(1+2k)\pi^2/{4}$, see Fig.~\ref{Fig7}~(b).
%
%The $\Delta B_{1}$ blows up at $NB_{1}T_{1}=k\pi$, since the value ${|\partial{\langle\hat{J}_{z}\rangle_{\text{f}}}/ \partial{B_{1}}|}$ approaches to $0$.
%
%The $\Delta B_{2}$ blows up at $NB_{2}T_{2}=(1+2k)\pi^2/{4}$, since the value ${|\partial{\langle\hat{J}_{z}\rangle_{\text{f}}}/ \partial{B_{2}}|}$ approaches to $0$.
%
The measurement precisions $\Delta B_{1}^{\textrm{sim}}$ and $\Delta B_{2}^{\textrm{sim}}$ both exhibit the Heisenberg-limited scalings with smaller constants simultaneously.
According to the fitting results, the optimal simultaneous measurement precisions $\Delta B_{1}^{\textrm{sim}}\approx \frac{\sqrt{2.5}}{{N}T_{1}}$ and $\Delta B_{2}^{\textrm{sim}}\approx \frac{\pi\sqrt{1.75}}{2{N}T_{2}}$, as shown in Fig.~\ref{Fig7}~(c) and (d) (stars).
%
%In other situations, the measurement precisions for estimating $B_{1}$ and $B_{2}$ can still preserve the Heisenberg-limited scaling simultaneously but with a constant.
%
This indicates that our scheme enables one to measure DC and AC magnetic fields approaching the Heisenberg limit simultaneously.
%
%%%%%%%%%%%%%%%%%%%%%%%%%%%%%%%%%%%%%%%%%%%%%%%%%%%%%%%%%%%%%%%%%%%%%%%%%%%%%%%%%%%%%%%%%%%%%%%%%%%%%%%%%%%%%%%%%%%%%%%%%%%%%%%%%%%%%%%%%%%%%%%%%%%%
%%%%%%%%%%%%%%%%%%%%%%%%%%%%%%%%%%%%%%%%%%%%%%%%%%%%%%%%%%%%%%%%%%%%%%%%%%%%%%%%%%%%%%%%%%%%%%%%%%%%%%%%%%%%%%%%%%%%%%%%%%%%%%%%%%%%%%%%
\section{Experimental feasibility\label{Sec4}}
%%%%%%%%%%%%%%%%%%%%%%%%%%%%%%%%%%%%%%%%%%%%%%%%%%%%%%%%%%%%%%%%%%%%%%%%%%%%%%%%%%%%%%%%%%%%%%%%%%%%%%%%%%%%%%%%%%%%%%%%%%%%%%%%%%%%%%%%
In this section, we will discuss the experimental feasibilities of our scheme.
To measure the DC and AC components simultaneously, the techniques of Ramsey interferometry and multiple rapid periodic $\pi$ pulses are needed in interrogation stage.
The precise implementation of $\pi/2$ and $\pi$ pulses is well-developed and widely used in synthetic quantum systems~\cite{GdeLange2010,PCMaurer2012,HStrobel2014,JGBohnet2016,ILovchinsky2016,SChoi2017,JMBossl2017,JRMaze2008,GBalasubramanian2008,GdeLange2011}. such as nuclear magnetic resonances, ultracold atoms and trapped ions.
For the second process of interrogation, the multi-$\pi$-pulse sequence is successionally applied with frequency $2\omega_s$, which can be implemented by the time sequence control system.

On the other hand, to further achieve the Heisenberg-limited simultaneous measurement, the generation of GHZ state and the implementation of interaction-based operation are the key processes for initialization and readout stages.
Owing to the controllable atom-atom interaction, it is feasible to realize the generation of GHZ state and the interaction-based operation via ultracold atoms.

As an example, we consider a cloud of trapped Bose condensed atoms occupying two hyperfine levels.
Under specific conditions, the system can be described by a symmetric two-mode Bose-Josephson Hamiltonian~\cite{HStrobel2014,Gross2010,Lee2006,Lee2009,Riedel2010,CLee2008,Ribeiro2007}, i.e.,
\begin{equation}\label{Eq:HamBJJ}
\hat{H}_{\text{BJ}}= -\Omega \hat J_x + \frac{\chi}{N} \hat J_z^2.
\end{equation}
The non-negative parameter $\Omega$ is the Josephson coupling strength, while $\chi$ denotes the nonlinear atom-atom interaction.
The atom-atom interaction is the key to produce entanglement among atoms.
The strength and the sign of $\chi$ can be tuned by modifying the s-wave scattering lengths via Feshbach resonance~\cite{HStrobel2014,Gross2010,Muessel2014} or adjusting the spatial overlap via spin-dependent forces~\cite{Riedel2010,Ockeloen2013}.

One way to generate the GHZ state is the dynamical evolution under the one-axis twisting Hamiltonian~\cite{Ferrini2010,Pawlowski2013,Spehner2014,Molmer1999,You2003,Linnemann2016,Lucke2011,Gross2010,Muessel2014,HStrobel2014}.
The one-axis twisting Hamiltonian $H_{\text{twist}}=\chi\hat{J}_z^2$ can be easily obtained by tuning the atom-atom interaction $\chi$ ($\chi>0$) dominant.
Initializing from a spin coherent state $\ket{\Psi}_{\textrm{SCS}}$, $H_{\text{twist}}$ generates non-Gaussian states,
including oversqueezed states and ultimately a maximally entangled GHZ state at $\chi t=\pi/2$.

Adiabatic evolution can also used for producing highly entangled states including GHZ state.
There exists a spontaneous symmetry-breaking transition between non-degenerate and degenerate groundstates when the atom-atom interaction is negative ($\chi<0$).
In strong coupling limit ($\Omega \gg |\chi|$), the system groundstate is approximately an SU(2) spin coherent state.
When $\Omega \rightarrow 0$, the atom-atom interaction dominates and the two lowest eigenstates become degenerate.
Due to parity symmetry, starting from the even-parity groundstate of $H_{BJ}$ with large $\Omega$ and adiabatically decreasing $\Omega$ across the critical point $\Omega_c= 1$, spin cat states (an even-parity eigenstate which is the superposition of two degenerate spin coherent states) can be produced.
Especially, when $\Omega$ decreases to $0$, one finally obtain a GHZ state~\cite{Lee2006, Lee2009,JHuang2015}.
This kind of symmetry-protected adiabatic evolution can be efficiently achieved by designing the time-dependent sweeping according to the symmetry-dependent adiabatic condition~\cite{MZhuang2018}.

The interaction-based operations $\hat{U}_{1}=e^{i\frac{\pi}{2}J_{z}^{2}}$ and $\hat{U}_{1}=e^{i\frac{\pi}{2}J_{x}^{2}}$ can also be realized by utilizing the atom-atom interaction.
Similar to the dynamical evolution for generating the GHZ state, the interaction-based operation $\hat{U}_{1}=e^{i\frac{\pi}{2}J_{z}^{2}}$ can be implemented via the anti-twisting process where only the sign of $\chi$ is changed~\cite{Linnemann2016,OHosten2016,Burd2019,OHosten2016}.
In addition, sandwiched a $-\pi/2$ pulse and a $\pi/2$ pulse (both along the y axis) between $\hat{U}_{1}=e^{i\frac{\pi}{2}J_{z}^{2}}$, the interaction-based operation $\hat{U}_{1}=e^{i\frac{\pi}{2}J_{x}^{2}}$ can also be achieved~\cite{Liu2011}.

%%%%%%%%%%%%%%%%%%%%%%%%%%%%%%%%%%%%%%%%%%%%%%%%%%%%%%%%%%%%%%%%%%%%%%%%%%%%%%%%%%%%%%%%%%%%%%%%%%%%%%%%%%%%%%%%%%%%%%%%%%%%%%%%%%%%%%%%
\section{summary \label{Sec5}}
%%%%%%%%%%%%%%%%%%%%%%%%%%%%%%%%%%%%%%%%%%%%%%%%%%%%%%%%%%%%%%%%%%%%%%%%%%%%%%%%%%%%%%%%%%%%%%%%%%%%%%%%%%%%%%%%%%%%%%%%%%%%%%%%%%%%%%%%
In summary, we have proposed a novel scheme for measuring DC and AC magnetic fields simultaneously.
In our scheme, the interrogation stage is divided into two signal accumulation processes and a unitary operation.
In the first process, regulating precisely the evolution time $T_{1}=n\frac{2\pi}{\omega_{s}}$, the effective Hamiltonian $\hat{H}_{B}^{\text{eff1}} = B_1 \hat J_z$ is only dependent on the DC component $B_{1}$.
In the second process, a multi-$\pi$-pulse sequence is applied for a duration $T_{2}$, in which the DC magnetic field make no contribution to the phase accumulation.
Thus, the effective Hamiltonian $\hat{H}_{B}^{\text{eff2}} = \frac{2B_2}{\pi} \hat J_z$ becomes only dependent on the AC component $B_{2}$.
Applying suitable unitary operations in interrogation and readout stages, one can extracting the DC and AC components $B_{1}$ and $B_{2}$ via population measurements.

Based upon the proposed scheme, we have studied the measurement precisions for simultaneous estimation of the DC and AC magnetic field with individual and entangled particles.
The detailed derivation of how the GHZ state can enhance the measurement precision is analytically given.
By employing the GHZ state as the input state and applying two suitable interaction-based operation, the measurement precisions of DC and AC magnetic fields may exhibit Heisenberg-limited scaling simultaneously.
The experimental feasibility of our scheme are also discussed.
This study highlights the multi-$\pi$-pulse sequence as a useful technique for quantum sensing of oscillating signals.
Our scheme may point out a new way for achieving Heisenberg-limited multiparameter estimation.

\acknowledgements{This work is supported by the Key-Area Research and Development Program of GuangDong Province under Grants No. 2019B030330001, the NSFC (Grant No. 11874434, No. 11574405, and No. 11704420), and the Science and Technology Program of Guangzhou (China) under Grants No. 201904020024.}

%%%%%%%%%%%%%%%%%%%%%%%%%%%%%%%%%%%%%%%%%%%%%%%%%%%%%%%%%%%%%%%%%%%%%%%%%%%%%%%%%%%%%%%%%%%%%%%%%%%%%%%%%%%%%%%%%%%%%%%%%%%%%%%%%%%%%%

\setcounter{equation}{0}
\renewcommand{\theequation}{A\arabic{equation}}
%%%%%%%%%%%%%%%%%%%%%%%%%%%%%%%%%%%%%%%%%%%%%%%%%%%%%%%%%%%%%%%%%%%%%%%%%%%%%%%%%%%%%%%%%%%%%%%%%%%%%%%%%%%%%%%%%%%%%%%%%%%%%%%%%%%%%%%%%%%%%%%
\section*{APPENDIX A}
%%%%%%%%%%%%%%%%%%%%%%%%%%%%%%%%%%%%%%%%%%%%%%%%%%%%%%%%%%%%%%%%%%%%%%%%%%%%%%%%%%%%%%%%%%%%%%%%%%%%%%%%%%%%%%%%%%%%%%%%%%%%%%%%%%%%%%%%%%%%%%%%
Here, we show how to calculate the QFIM and CFIM in detail.
The QFIM determines the ultimate precision bound for simultaneous multiparameter measurement, and it can be derived according to the output state after interrogation.
The elements of the QFIM can be written as
\begin{equation}\label{Eq:FQ}
 \left[\mathnormal{\mathbf{F}}_Q(B_{1},B_{2})\right]_{kl}=\mathrm{Tr}\left[\rho_{\textrm{Out}}\frac{\hat{L}_k\hat{L}_l+\hat{L}_l\hat{L}_k}{2}\right],
\end{equation}
where $k,l=1,2$, $\hat{L}_k$ is symmetric logarithmic derivative (SLD) and $\rho_{\textrm{Out}}=\ket{\Psi_{\textrm{Out}}}\bra{\Psi_{\textrm{Out}}}$ is the density matrix of the output state.

Within our scheme described in Sec.~\ref{Sec2} of the main text, the SLD can be explicitly expressed as
\begin{eqnarray}\label{Eq:SLD_q}
  \hat{L}_k=2\left(\ket{\partial_{B_{k}}\Psi_{\textrm{Out}}}\bra{\Psi{_{\textrm{Out}}}}\!+\!\ket{\Psi_{\textrm{Out}}}\bra{\partial_{B_{k}}\Psi_{\textrm{Out}}}\right),
\end{eqnarray}
where $k=1,2$ and $\ket{\partial_{B_{k}}\Psi_{\textrm{Out}}}$ denotes the partial derivative of $\ket{\Psi_{\textrm{Out}}}$ with respect to the parameter $B_{k}$.
The elements of the QFIM can be further simplified as
\begin{eqnarray}\label{Eq:FQM}
  \left[\mathnormal{\mathbf{F}}_Q(B_{1},B_{2})\right]_{kl}= && 4\bra{\Psi_{\textrm{In}}}e^{i B_{1} \hat{J}_{z} T_{1}} \hat{H}_{k}\hat{H}_{l}e^{-i B_{1} \hat{J}_{z} T_{1}} \ket{\Psi_{\textrm{In}}}\nonumber\\
  && -4 \bra{\Psi_{\textrm{In}}}e^{i B_{1} \hat{J}_{z} T_{1}} \hat{H}_{k} e^{-i B_{1} \hat{J}_{z} T_{1}}\ket{\Psi_{\textrm{In}}}\nonumber\\
  && \bra{\Psi_{\textrm{In}}}e^{i B_{1} \hat{J}_{z} T_{1}} \hat{H}_{l} e^{-i B_{1} \hat{J}_{z} T_{1}} \ket{\Psi_{\textrm{In}}},
\end{eqnarray}
where $k,l=1,2$ and $\hat{H}_{1}=\hat{J}_{z}T_{1}$, $\hat{H}_{2}=\hat{U}_{1}^{\dag}(2\hat{J}_{z}T_{2}/\pi) \hat{U}_{1}$.

To calculate the CFIM, one needs to choose a certain measurement that is generally described by POVM.
POVM can be defined as a collection $\{\hat{\Pi}_m\}$ of positive operators on a Hilbert space that sum to the identity,
\begin{equation}\label{POVM}
    \sum_m \hat{\Pi}_m =\hat{I},
\end{equation}
and $\bra{\psi}\hat{\Pi}_m\ket{\psi} \ge 0$ for any $\ket{\psi}$.
Here, $m$ stands for the outcome of the measurement.

Within our scheme described in Sec.~\ref{Sec2} of the main text, for the final state, the conditional probability with outcome $m$ given the parameters $B_1$ and $B_2$ is expressed as
\begin{equation}\label{ConditionalPro}
    P_m(B_1,B_2)=\bra{\Psi_{\text{final}}}\hat{\Pi}_m\ket{\Psi_{\text{final}}},
\end{equation}
where $\ket{\Psi_{\text{final}}}$ contains the information of parameters $B_1$ and $B_2$.
Given the conditional probability $P_m(B_1,B_2)$, the CFIM can be obtained.
The element of the CFIM reads
\begin{eqnarray}\label{Eq:FIM}
&&\left[\mathnormal{\mathbf{F}}_C(B_{1},B_{2})\right]_{kl}\nonumber\\
&&=\sum_{m}\frac{1}{P_{m}(B_1,B_2)}\frac{\partial{P_{m}(B_1,B_2)}}{\partial{B_{k}}} \frac{\partial{P_{m}(B_1,B_2)}}{\partial{B_{l}}}.
\end{eqnarray}

Provided the QFIM and CFIM, the QCRB and the CCRB can be easily obtained according to Eq.~\eqref{InEq:Delta_B1_B2_general}. In Sec.~\ref{Sec3}, we have analytically give the QCRB and the CCRB for the three specific cases and numerically illustrate the CCRB in Fig.~\ref{Fig5}.

%
%%%%%%%%%%%%%%%%%%%%%%%%%%%%%%%%%%%%%%%%%%%%%%%%%%%%%%%%%%%%%%%%%%%%%%%%%%%%%%%%%%%%%%%%%%%%%%%%%%%%%%%%%%%%%%%%%%%%%%%%%%%%%%%%%%%%%%%%%%%%%%%
\section*{APPENDIX B}
%%%%%%%%%%%%%%%%%%%%%%%%%%%%%%%%%%%%%%%%%%%%%%%%%%%%%%%%%%%%%%%%%%%%%%%%%%%%%%%%%%%%%%%%%%%%%%%%%%%%%%%%%%%%%%%%%%%%%%%%%%%%%%%%%%%%%%%%%%%%%%%%
Here, we give the the proof of Eq.\eqref{Evo_CSC} in the main text and we choose the Dike basis satisfy: $\hat{J}_{z}|J,m_z\rangle=m_z|J,m_z\rangle$.
%
%In our calculation, the durations $T_1$ and $T_2$ are set to be the same, i.e., $T_1=T_2=T$.
%
Thus,
\begin{eqnarray}\label{Eq:All_down}
|{J,-J}\rangle=\prod^{2J}_{l=1}|\downarrow\rangle_{l} \quad;\quad |{J,J}\rangle=\prod^{2J}_{l=1}|\uparrow\rangle_{l}.
\end{eqnarray}
%
%\begin{eqnarray}\label{Eq:All_up}
%|{J,J}\rangle=\prod^{2J}_{l=1}|\uparrow\rangle
%\end{eqnarray}
%
First,
\begin{eqnarray}\label{Eq:Evo_CSC_1}
&&e^{-i\frac{\pi}{2}{\hat{J}_{y}}}e^{-i\hat{{H}}_{B}^{\text{eff1}}T_{1}}e^{-i\frac{\pi}{2}{\hat{J}_{y}}}|{J,-J}\rangle \nonumber \\
&&= e^{-i\frac{\pi}{4}\sum^{2J}_{l=1}\hat{\sigma}_{y}}e^{{-\frac{iB_1T_{1}}{2}\sum^{2J}_{l=1}\hat{\sigma}_{z}}} e^{-i\frac{\pi}{4}\sum^{2J}_{l=1}\hat{\sigma}_{y}} \prod^{2J}_{l=1}|\downarrow\rangle_{l} \nonumber \\
&&=e^{-i\frac{\pi}{4}\sum^{2J}_{l=1}\hat{\sigma}_{y}}e^{{-\frac{iB_1T_{1}}{2}\sum^{2J}_{l=1}\hat{\sigma}_{z}}}\prod^{2J}_{l=1}(-\frac{1}{\sqrt{2}}|\uparrow\rangle_{l}+\frac{1}{\sqrt{2}}|\downarrow\rangle_{l})
\nonumber \\
&&=e^{-i\frac{\pi}{4}\sum^{2J}_{l=1}\hat{\sigma}_{y}}\prod^{2J}_{l=1}\frac{1}{\sqrt{2}}(-e^{-i{{B_1T_{1}/2}}}|\!\uparrow\rangle_{l}+e^{i{B_1T_{1}/2}}|\!\downarrow\rangle_{l})
\nonumber \\
&&=\prod^{2J}_{l=1}\frac{1}{2}\left[-e^{-i{B_1T_{1}/2}}(|\!\uparrow\rangle_{l}\!+|\!\downarrow\rangle_{l})+e^{i{B_1T_{1}/2}}(-|\!\uparrow\rangle_{l}\!+|\!\downarrow\rangle_{l})\right] \nonumber \\
&&= \prod^{2J}_{l=1}\left[-\cos({B_1T_{1}/2})|\!\uparrow\rangle_{l} + i\sin({B_1T_{1}/2})|\!\downarrow\rangle_{l}\right].
\end{eqnarray}

Then,
\begin{eqnarray}\label{Eq:Evo_CSC_2}
|\Psi\rangle_\textrm{final}&=&e^{-i\frac{\pi}{2}{\hat{J}_{y}}}e^{-i\hat{H}_{B}^{\text{eff2}}T_{2}}e^{-i\frac{\pi}{2}{\hat{J}_{y}}}e^{-i\hat{H}_{B}^{\text{eff1}}T_{1}}e^{-i\frac{\pi}{2}{\hat{J}_{y}}}|{J,-J}\rangle \nonumber \\
&=&e^{-\frac{i\pi}{4}\sum^{2J}_{l=1}\hat{\sigma}_{y}}e^{-i\frac{B_2T_{2}}{{\pi}}\sum^{2J}_{l=1}\hat{\sigma}_{z}}\!\!\prod^{2J}_{l=1}\!\left[-\cos({B_1T_{1}/2})|\uparrow\rangle_{l}\right]\nonumber\\
&\quad&+e^{-\frac{i\pi}{4}\sum^{2J}_{l=1}\hat{\sigma}_{y}}e^{-i\frac{B_2T_{2}}{\pi}\sum^{2J}_{l=1}\hat{\sigma}_{z}}\prod^{2J}_{l=1} \left[i\sin({B_1T_{1}/2})|\downarrow\rangle_{l}\right]\nonumber \\
&=&e^{-\frac{i\pi}{4}\!\!\sum^{2J}_{l=1}\hat{\sigma}_{y}}\prod^{2J}_{l=1}\left[-\cos({B_1T_{1}/2})e^{-{i{B_2T_{2}}}/{\pi}}|\!\uparrow\rangle_{l}\right]
\nonumber \\
&\quad&+e^{-i\frac{\pi}{4}\sum^{2J}_{l=1}\hat{\sigma}_{y}}\prod^{2J}_{l=1}\left[i\sin({B_1T_{1}/2})e^{{i{B_2T_{2}}}/{\pi}}|\downarrow\rangle_{l}\right]
\nonumber \\
&=&\prod^{2J}_{l=1}\left[-\cos({B_1T_{1}/2})e^{-i{B_2T_{2}/{\pi}}}(\frac{1}{\sqrt{2}}|\!\uparrow\rangle_{l}+\frac{1}{\sqrt{2}}|\!\downarrow\rangle_{l})\right]
 \nonumber \\
&\quad&+\prod^{2J}_{l=1}\left[i\sin({B_1T_{1}/2})e^{i{B_2T_{2}/{\pi}}}(-\frac{1}{\sqrt{2}}|\!\uparrow\rangle_{l}+\frac{1}{\sqrt{2}}|\!\downarrow\rangle_{l})\right]
\nonumber \\
&=&\frac{1}{\sqrt{2}}\!\prod^{2J}_{l=1}\!\left[-\cos(\frac{B_1T_{1}}{2})e^{-i\frac{B_2T_{2}}{\pi}}\!\!\!-\!\!i\sin(\frac{B_1T_{1}}{2})e^{i\frac{B_2T_{2}}{\pi}}\right]|\!\uparrow\rangle_{l}
 \nonumber \\
&\quad&+\frac{1}{\sqrt{2}}\!\prod^{2J}_{l=1}\!\left[-\cos(\frac{B_1T_{1}}{2})e^{-i\frac{B_2T_{2}}{\pi}}\!\!\!+\!\!i\sin(\frac{B_1T_{1}}{2})e^{i\frac{B_2T_{2}}{\pi}}\right]\!\!|\!\downarrow\rangle_{l}.
\nonumber \\
\end{eqnarray}
So, we can unify Eq.~\eqref{Eq:Evo_CSC_2} as Eq.~\eqref{Evo_CSC} in the main text.
%

%%%%%%%%%%%%%%%%%%%%%%%%%%%%%%%%%%%%%%%%%%%%%%%%%%%%%%%%%%%%%%%%%%%%%%%%%%%%%%%%%%%%%%%%%%%%%%%%%%%%%%%%%%%%%%%%%%%%%%%%%%%%%%%%%%%%%%%%%%%%%%%
\section*{APPENDIX C}
%%%%%%%%%%%%%%%%%%%%%%%%%%%%%%%%%%%%%%%%%%%%%%%%%%%%%%%%%%%%%%%%%%%%%%%%%%%%%%%%%%%%%%%%%%%%%%%%%%%%%%%%%%%%%%%%%%%%%%%%%%%%%%%%%%%%%%%%%%%%%%%%
For simplicity, we consider $J=N/2$ is even number in here and give the proof of Eq.\eqref{Evo_GHZ1_even} in the main text.
The proof of Eq.\eqref{Evo_GHZ1_odd} in the main text also can be obtained according to this section.
First, in the Dike basis, the GHZ state is

\begin{eqnarray}\label{Eq:GHZ}
|\Psi\rangle_{\text{GHZ}}=\frac{1}{\sqrt{2}}(|{J,-J}\rangle+|{J,J}\rangle)=\frac{1}{\sqrt{2}}(\prod^{2J}_{l=1}|\!\downarrow\rangle_{l}+\prod^{2J}_{l=1}|\!\uparrow\rangle_{l}).\nonumber\\
\end{eqnarray}
Thus,
\begin{eqnarray}\label{Eq:GHZ0}
&e&^{-i\hat{H}_{B}^{\text{eff1}}T_{1}}|\Psi\rangle_{\text{GHZ}}\nonumber\\
&=&e^{-{iB_1T_{1}}/{2}\sum^{2J}_{l=1}\hat{\sigma}_{z}}\frac{1}{\sqrt{2}}
(\prod^{2J}_{l=1}|\!\downarrow\rangle_{l}+\prod^{2J}_{l=1}|\!\uparrow\rangle_{l}) \nonumber\\
&=&\frac{1}{\sqrt{2}}(\prod^{2J}_{l=1}e^{-i{B_1T_{1}/2}}|\!\uparrow\rangle_{l}+\prod^{2J}_{l=1}e^{i{B_1T_{1}/2}}|\!\downarrow\rangle_{l}),\nonumber\\
\end{eqnarray}
 $\quad$$\quad$$\quad$ and
\begin{widetext}
\begin{eqnarray}\label{Eq:GHZ1}
e^{-i\frac{\pi}{2}{\hat{J}_{y}}}e^{-i\hat{H}_{B}^{\text{eff1}}T_{1}}|\Psi\rangle_{\text{GHZ}}
&=&e^{-i\frac{\pi}{4}\sum^{2J}_{l=1}\hat{\sigma}_{y}}\frac{1}{\sqrt{2}}(\prod^{2J}_{l=1}e^{-i{B_1T_{1}/2}}|\!\uparrow\rangle_{l}+\prod^{2J}_{l=1}e^{i{B_1T_{1}/2}}|\!\downarrow\rangle_{l}) \nonumber\\
&=&\frac{1}{{2}}e^{-i{B_1JT_{1}}}\prod^{2J}_{l=1}(|\!\uparrow\rangle_{l}+|\!\downarrow\rangle_{l})
+\frac{1}{{2}}e^{i{B_1JT_{1}}}\prod^{2J}_{l=1}(-|\!\uparrow\rangle_{l}+|\!\downarrow\rangle_{l}) \nonumber\\
&=&\!\frac{1}{\sqrt{2}}e^{-i{B_1JT_{1}}}\!\!\sum^{J}_{m_z=-J}\!\frac{C^{m_z}_{J}}{2^{J}} |{J,m_z}\rangle
+\frac{1}{\sqrt{2}}e^{i{B_1JT_{1}}}\!\!\sum^{J}_{m_z=-J}\!\frac{C^{m_z}_{J}}{2^{J}}(-1)^{J+m_z}|{J,m_z}\rangle.
\end{eqnarray}

Here, $C_{J}^{m_z}={\frac{(2J)!}{(J+m_z)!(J-m_z)!}}$ is the binomial coefficient. Then,

\begin{eqnarray}\label{Eq:GHZ2}
&e&^{i\frac{\pi}{2}{\hat{J}_{z}^{2}}}e^{-i\hat{H}_{B}^{\text{eff2}}T_{2}}e^{-i\frac{\pi}{2}{\hat{J}_{y}}}e^{-i\hat{H}_{B}^{\text{eff1}}T_{1}}|\Psi\rangle_{\text{GHZ}} \nonumber\\
&=&\frac{1}{\sqrt{2}}e^{-i{B_1JT_{1}}}e^{i\frac{\pi}{2}{\hat{J}_{z}^{2}}}e^{-i{(2B_2 T_{2}/\pi)}\hat{J}_{z}}\!\sum^{J}_{m_z=-J}\!\frac{C^{m_z}_{J}}{2^{J}}|{J,m_z}\rangle
%\nonumber\\
+\frac{1}{\sqrt{2}}e^{i{B_1JT_{1}}}e^{i\frac{\pi}{2}{\hat{J}_{z}^{2}}}e^{-i{(2B_2T_{2}/\pi)}\hat{J}_{z}}\!\sum^{J}_{m_z=-J}\!\frac{C^{m_z}_{J}}{2^{J}}(-1)^{J+m_z}|{J,m_z}\rangle
\nonumber\\
&=&\frac{1}{\sqrt{2}}e^{-i{B_1JT_{1}}}\!\sum^{J}_{m_z=-J}\!\frac{C^{m_z}_{J}}{2^{J}}(i)^{m_z^2}e^{-i(2{B_2}m_z T_{2}/{\pi})}|{J,m_z}\rangle
 %\nonumber\\
+\frac{1}{\sqrt{2}}e^{i{B_1JT_{1}}}\!\sum^{J}_{m_z=-J}\!\frac{C^{m_z}_{J}}{2^{J}}(-1)^{J+m_z}(i)^{m_z^2}e^{-i(2{B_2}m_z T_{2}/{\pi})}|{J,m_z}\rangle. \nonumber\\
\end{eqnarray}
Considering the cases of even and odd $m$ respectively, we surprisingly find that,
\begin{eqnarray}\label{Eq:GHZ3}
(i)^{m^2}\left[e^{-i{B_1JT_{1}}}\!+\!(-\!1)^{J+m}e^{i{B_1JT_{1}}}\right]= \cos(B_1JT_{1})\left[(-1)^{m}+1\right]-\sin(B_1JT_{1})\left[(-1)^{m}-1\right]. \nonumber\\
\end{eqnarray}

Substituting Eq.\eqref{Eq:GHZ3} into Eq.\eqref{Eq:GHZ2}, we find that
\begin{eqnarray}\label{Eq:GHZ4}
|\Psi\rangle_{\text{final}}&=& e^{-i\frac{\pi}{2}{\hat{J}_{y}}}e^{i\frac{\pi}{2}{\hat{J}_{z}^{2}}}e^{-i\hat{H}_{B_2}^{\text{eff2}}T_{2}}e^{-i\frac{\pi}{2}{\hat{J}_{y}}}e^{-i\hat{H}_{B_1}^{\text{eff1}}T_{1}} |\Psi\rangle_{\text{GHZ}} \nonumber\\
&=&\frac{e^{-i\frac{\pi}{2}{\hat{J}_{y}}}}{\sqrt{2}}\sum^{J}_{m_z=-J} \frac{C^{m_z}_{J}}{2^{J}} e^{-i(\frac{2{B_2}m_z T_{2}}{\pi})}[\cos(B_1JT_{1})+ \sin(B_1JT_{1})]|{J,m_z}\rangle \nonumber\\
&&+\frac{e^{-i\frac{\pi}{2}{\hat{J}_{y}}}}{2}\sum^{J}_{m_z=-J}\frac{C^{m_z}_{J}}{2^{J}}e^{-i(\frac{2{B_2}m_z T_{2}}{\pi})}[\cos(B_1JT_{1})-\sin(B_1JT_{1})] (-1)^{m_z}|{J,m_z}\rangle\ .
\end{eqnarray}
According to Eqs.~\eqref{Eq:GHZ3} and \eqref{Eq:GHZ4},
\begin{eqnarray}\label{Eq:GHZ5}
\sum^{J}_{m_z=-J} \frac{C^{m_z}_{J}}{2^{J}}e^{-i\frac{2{B_2}m_z T_{1}}{\pi}}(-1)^{m_z}|{J,m_z}\rangle
=\prod^{2J}_{l=1}\frac{i}{\sqrt{2}}(e^{-i\frac{2B_2T_{2}}{\pi}}|\uparrow\rangle_{l}+ e^{i\frac{2B_2 T_{2}}{\pi}}|\downarrow\rangle_{l}),
\end{eqnarray}
\begin{eqnarray}\label{Eq:GHZ6}
\sum^{J}_{m_z=-J} \frac{C^{m_z}_{J}}{2^{J}}e^{-i\frac{{2B_2}m_z T_{2}}{\pi}}|{J,m_z}\rangle
=\prod^{2J}_{l=1}\frac{1}{\sqrt{2}}(e^{-i\frac{2B_2 T_{2}}{\pi}}|\uparrow\rangle_{l}+e^{i\frac{2B_2 T_{2}}{\pi}}|\downarrow\rangle_{l}).
\end{eqnarray}
Substituting Eq.~\eqref{Eq:GHZ5} and Eq.~\eqref{Eq:GHZ6} into Eq.~\eqref{Eq:GHZ4}, the final state becomes
\begin{eqnarray}\label{Eq:GHZ7}
|\Psi\rangle_{\text{final}}&=& e^{-i\frac{\pi}{2}{\hat{J}_{y}}}e^{i\frac{\pi}{2}{\hat{J}_{z}^{2}}}e^{-i\hat{H}_{B}^{\text{eff2}} T_{2}}e^{-i\frac{\pi}{2}{\hat{J}_{y}}}e^{-i\hat{H}_{B}^{\text{eff1}} T_{1}}|\Psi\rangle_{\text{GHZ}}\nonumber\\
&=& e^{-i\frac{\pi}{2}{\hat{J}_{y}}} [\cos(B_1JT) - \sin(B_1JT_{1})] \prod^{2J}_{l=1}\frac{1}{{2}}\left(i e^{-i{\frac{2B_2 T_{2}}{\pi}}}|\uparrow\rangle_{l}+ e^{i\frac{2B_2 T_{2}}{\pi}} |\downarrow\rangle_{l}\right)\nonumber\\
&\quad&+ e^{-i\frac{\pi}{2}{\hat{J}_{y}}}[\cos(B_1JT_{1}) +\sin(B_1JT_{1})] \prod^{2J}_{l=1}\frac{1}{2}\left(e^{\frac{-iB_2T_{2}}{\pi}}|\uparrow\rangle_{l} + e^{-i{\frac{2B_2T_{2}}{\pi}}}|\downarrow\rangle_{l}\right) \nonumber\\
&=& \frac{1}{\sqrt{2}} [\cos(B_1JT) -\sin(B_1JT_{1})] \prod^{2J}_{l=1} \left[i\cos(\frac{{2B_2T_{2}}}{\pi})|\uparrow\rangle_{l} +\sin(\frac{{2B_2T_{2}}}{\pi})|\downarrow\rangle_{l}\right] \nonumber\\
&\quad&+ \frac{1}{\sqrt{2}}[\cos(B_1JT) +\sin(B_1JT_{1})] \prod^{2J}_{l=1} \left[-i\sin(\frac{{2B_2T_{2}}}{\pi})|\uparrow\rangle_{l} +\cos(\frac{{2B_2T_{2}}}{\pi})|\downarrow\rangle_{l})\right].
\end{eqnarray}
Finally, we can unify Eq.~\eqref{Eq:GHZ7} as Eq.~\eqref{Evo_GHZ1_even} in the main text.

%
%Now, we have given the proof of E.q.~\eqref{Evo_GHZ1_even} in the main text.
%%%%%%%%%%%%%%%%%%%%%%%%%%%%%%%%%%%%%%%%%%%%%%%%%%%%%%%%%%%%%%%%%%%%%%%%%%%%%%%%%%%%%%%%%%%%%%%%%%%%%%%%%%%%%%%%%%%%%%%%%%%%%%%%%%%%%%%%%%%%%%%
\section*{APPENDIX D}
%%%%%%%%%%%%%%%%%%%%%%%%%%%%%%%%%%%%%%%%%%%%%%%%%%%%%%%%%%%%%%%%%%%%%%%%%%%%%%%%%%%%%%%%%%%%%%%%%%%%%%%%%%%%%%%%%%%%%%%%%%%%%%%%%%%%%%%%%%%%%%%%
In this section, we will give the proof of Eq.~\eqref{Evo_GHZ2} in the main text.
For simplicity, we only consider $J=N/2$ is even in here.
%
%First, we need to introduce the effective of unitary operator $e^{\frac{i\pi}{2}\hat{S}_{x}^{2}}$ on two
According to the Eq.~\eqref{Eq:GHZ1}, we have
\begin{equation}\label{Eq:GHZ8}
e^{i\frac{\pi}{2}{\hat{J}_{x}}^2}e^{-i\hat{H}_{B}^{\text{eff1}}T_{1}}|\Psi\rangle_{\text{GHZ}} = e^{i\frac{\pi}{2}{\hat{J}_{x}}^2}\left(\frac{1}{\sqrt{2}} e^{-i{B_1JT_{1}}}|J,J\rangle + \frac{1}{\sqrt{2}} e^{i{B_1JT_{1}}} |J,-J\rangle\right).
\end{equation}
Since
\begin{equation}\label{Eq:Sx2_1}
e^{i\frac{\pi}{2}{\hat{J}_{x}}^2}|J,J\rangle=e^{i(\pi/4+N\pi/2)}|J,J\rangle+e^{-i\pi/4}|J,-J\rangle,
\end{equation}
and
\begin{equation}\label{Eq:Sx2_2}
e^{i\frac{\pi}{2}{\hat{J}_{x}}^2}|J,-J\rangle=e^{-i(\pi/4+N\pi/2)}|J,J\rangle+e^{i\pi/4}|J,-J\rangle,
\end{equation}
we have
\begin{equation}\label{Eq:GHZ9}
e^{i\frac{\pi}{2}{\hat{J}_{x}}^2}e^{-i\hat{H}_{B}^{\text{eff1}}T_{1}}|\Psi\rangle_{\text{GHZ}} =\cos[-B_1JT_{1}+(\pi/4+{N\pi}/{2})]|J,J\rangle+\cos(B_1JT_{1}+{\pi}/{4})|J,-J\rangle.
\end{equation}
Similar to the procedures from Eq.~\eqref{Eq:GHZ2} to Eq.~\eqref{Eq:GHZ4}, we have
\begin{eqnarray}\label{Eq:GHZ9}
&&|\Psi\rangle_{\text{final}} = e^{i\frac{\pi}{2}{\hat{J}_{z}}^2}e^{i\frac{\pi}{2}{\hat{J}_{y}}}e^{-i\hat{H}_{B}^{\text{eff2}}T_{2}} e^{i\frac{\pi}{2}{\hat{J}_{x}}^2} e^{-i\hat{H}_{B}^{\text{eff1}}T} |\Psi\rangle_{\text{GHZ}} \nonumber \\
&=& \frac{1}{\sqrt{2}}[\cos(B_1JT_{1})\cos({B_2 J T_{2}}/{\pi}) -i \sin(B_1 J T_{1})\sin({B_2 J T_{2}}/{\pi})]\sum^{J}_{m_z=-J} \frac{C^{m_z}_{J}}{2^{J}}[(-1)^{m_z}+1]|J,m_z\rangle \nonumber \\
&\quad&-\frac{1}{\sqrt{2}}[\cos(B_1JT_{1})\sin({B_2 J T_{2}}/{\pi}) +i \sin(B_1JT)\cos({B_2 J T_{2}}/{\pi})]\sum^{J}_{m_z=-J} \frac{C^{m_z}_{J}}{2^{J}}[(-1)^{m_z}-1]|J,m_z\rangle \nonumber \\
&=& \frac{1}{\sqrt{2}}\{\cos(B_1JT_{1})[\cos({B_2 J T_{2}}/{\pi})-\sin({B_2JT_{2}}/{\pi})] -i \sin(B_1J T_{1})[\cos({B_2JT_{2}}/{\pi})+\sin({B_2 J T_{2}}/{\pi})]\}\!\!\sum^{J}_{m_z=-J}\frac{C^{m_z}_{J}}{2^{J}}(-1)^{m_z}|J,m_z\rangle  \nonumber\\
&\quad&+ \frac{1}{\sqrt{2}}\{\cos(B_1JT_{1})[\cos({B_2JT_{2}}/{\pi})+\sin({B_2JT}/{\pi})] +i \sin(B_1JT)[\cos({B_2JT_{2}}/{\pi})-\sin({B_2JT_{2}}/{\pi})]\}\!\sum^{J}_{m_z=-J}\frac{C^{m_z}_{J}}{2^{J}}|J,m_z\rangle . \nonumber\\
\end{eqnarray}
Finally, combining Eq.~\eqref{Eq:GHZ5} and Eq.~\eqref{Eq:GHZ6}, we can obtain the final state Eq.~\eqref{Evo_GHZ2} in the main text.
\end{widetext}
%%%%%%%%%%%%%%%%%%%%%%%%%%%%%%%%%%%%%%%%%%%%%%%%%%%%%%%%%%%%%%%%%%%%%%%%%%%%%%%%%%%%%%%%%%%

\end{document}